\documentclass[twocolumn,showpacs,superscriptaddress,prx,citeautoscript,amsmath,amssymb,letterpaper]{revtex4-1}

\usepackage{amsfonts}
\usepackage{amsthm}
\usepackage{graphicx}
\usepackage{multirow}
\usepackage{color}
\usepackage{bbold}
\usepackage{bm}
\usepackage{times}
\usepackage{amsmath,bm,amsfonts}
\usepackage{dcolumn}
\usepackage{graphicx}
\usepackage{latexsym}

\usepackage{ulem} 
\usepackage{braket}

\begin{document}

\title{Fractional charge bound to a vortex in two-dimensional topological crystalline insulators}

\author{Eunwoo \surname{Lee}}
\affiliation{Department of Physics and Astronomy, Seoul National University, Seoul 08826, Korea}

\affiliation{Center for Correlated Electron Systems, Institute for Basic Science (IBS), Seoul 08826, Korea}

\affiliation{Center for Theoretical Physics (CTP), Seoul National University, Seoul 08826, Korea}

\author{Akira \surname{Furusaki}}
\affiliation{RIKEN Center for Emergent Matter Science, Wako, Saitama, 351-0198, Japan}

\affiliation{Condensed Matter Theory Laboratory, RIKEN, Wako, Saitama, 351-0198, Japan}

\author{Bohm-Jung \surname{Yang}}
\email{bjyang@snu.ac.kr}
\affiliation{Department of Physics and Astronomy, Seoul National University, Seoul 08826, Korea}

\affiliation{Center for Correlated Electron Systems, Institute for Basic Science (IBS), Seoul 08826, Korea}

\affiliation{Center for Theoretical Physics (CTP), Seoul National University, Seoul 08826, Korea}

\date{\today}

\begin{abstract}
We establish the correspondence between the fractional charge bound to a vortex in a textured lattice and the relevant bulk band topology in two-dimensional (2D) topological crystalline insulators.
As a representative example, we consider the Kekule textured graphene whose bulk band topology is characterized by a 2D $\mathbb{Z}_{2}$ topological invariant $\nu_{\rm 2D}$ protected by inversion symmetry.
The fractional charge localized at a vortex in the Kekule texture is shown to be related to the change in the bulk topological invariant $\nu_{\rm 2D}$ around the vortex, as in the case of the Su-Schriefer-Heeger model in which the fractional charge localized at a domain wall is related to the change in the bulk charge polarization between degenerate ground states. We show that the effective three-dimensional (3D) Hamiltonian, where the angle $\theta$ around a vortex in Kekule-textured graphene is a third coordinate, describes a 3D axion insulator with a quantized magnetoelectric polarization. The spectral flow during the adiabatic variation of $\theta$ corresponds to the chiral hinge modes of an axion insulator and determines the accumulated charge localized at the vortex, which is half-quantized when chiral symmetry exists. When chiral symmetry is absent, electric charge localized at the vortex is no longer quantized, but the vortex always carries a half-quantized Wannier charge as long as inversion symmetry exists. For the cases when magnetoelectric polarization is quantized due to the presence of symmetry that reverses the space-time orientation, we classify all possible topological crystalline insulators whose vortex defect carries a fractional charge. 
\end{abstract}


\maketitle

\section{Introduction}
Fractional charge localized at topological defects is closely related to the topology of bulk electronic states~\cite{su1980soliton, jackiw1976solitons, jackiw1981zero, goldstone1981fractional, read2000paired, teo2010majorana, teo2010topological}.
For instance, the one-dimensional (1D) model proposed by Su, Schrieffer, and Heeger (SSH) is a representative system where a half electric charge is localized at a domain wall~\cite{su1980soliton}. Spontaneous formation of lattice dimerization gives rise to two degenerate ground states, and the zero-mode carrying the half electric charge is localized at the domain wall interpolating between the degenerate ground states. The inherent relationship between the zero-mode charge and the topology of the bulk electronic states can be seen from the quantized charge polarization $P_1$ of the two degenerate ground states, given by $P_1=0$ and $P_1=1/2$, respectively. The fractional charge accumulated at the domain wall is determined by the difference of the bulk charge polarization $P_1$ of the two ground states.

A remarkable idea realizing fractional charge in two dimensions is proposed by Hou, Chamon, and Mudry (HCM) in Ref.~[\onlinecite{hou2007electron}]. They considered a graphenelike system, whose low-energy excitations are described by massless Dirac fermions, and showed that a vortex in the order parameter for the Kekule-type dimerization accommodates zero-energy bound states with fractional charge. Spontaneous formation of the Kekule texture in graphene leads to the degenerate ground states with broken lattice symmetries~\cite{hou2007electron, chamon2008electron,chamon2008irrational, ryu2009masses}. In view of the connection between the charge fractionalization and the topological properties (quantized polarization) of the degenerate ground states in the SSH model, the charge fractionalization found in the HCM model naturally leads to the following question: Is it possible to understand the fractional charge at the vortex from topological properties of the bulk electronic states in two dimensions? 

In this paper, we show that the ground states of Kekule textured graphene (KTG) are characterized by a $\mathbb{Z}_{2}$ topological invariant $\nu_{\rm 2D}$, which is quantized in two-dimensional (2D) systems with inversion symmetry $P$~\cite{hughes2011inversion, morimoto2014weyl, qi2008topological}. Two insulating ground states with distinct quantized values of $\nu_{\rm 2D}$ can be changed by switching the strong and weak bonds in the Kekule texture, as in the case of the SSH model where two gapped phases with distinct $P_1$ can be interchanged in an analogous way. 
We show that KTG with $\nu_{\rm 2D}=1$ is a 2D inversion-symmetric second-order topological insulator (SOTI) with zero-energy corner states.
The nontrivial bulk band topology of KTG is confirmed by calculating the parity eigenvalues, the Wilson loop spectra, and the corner charge distribution. 
We note that the higher-order band topology of KTG related with sixfold rotation symmetry was discussed in Refs.~\cite{noh2018topological, benalcazar2018quantization}.
Similar to the case of SSH model, the charge accumulation at a vortex in the order parameter in the HCM model is related to the difference of the bulk topological quantity $\nu_{\rm 2D}$ between the two degenerate ground states with $\nu_{\rm 2D}=0$ and $\nu_{\rm 2D}=1$, respectively. 
Furthermore, we demonstrate that the change in $\nu_{\rm 2D}$ around a vortex is related to the quantized magnetoelectric polarizability $P_3$ of an axion insulator, which guarantees the presence of a zero-mode state localized at the vortex core.

Finally, considering that $P_3$ is quantized in the presence of the symmetry reversing the space-time orientation, we classify all possible topological crystalline insulators where an order parameter vortex can support fractional vortex charge. It is worth noting that the correspondence between the bulk band topology and the vortex bound state in KTG is beyond the ten-fold classification scheme of defect states proposed by Teo and Kane~\cite{teo2010topological}, in which crystalline symmetries are not considered.
We note that the electric charge localized at the vortex is half-quantized only when chiral symmetry exists, similar to the cases considered by Teo and Kane~\cite{teo2010topological}.
However, due to the quantized $P_3$ protected by inversion symmetry, the vortex always carries a quantized Wannier charge even if the electric charge accumulated at the vortex is not quantized when chiral symmetry is broken.

\begin{figure}[t]
\includegraphics[width=8.5 cm]{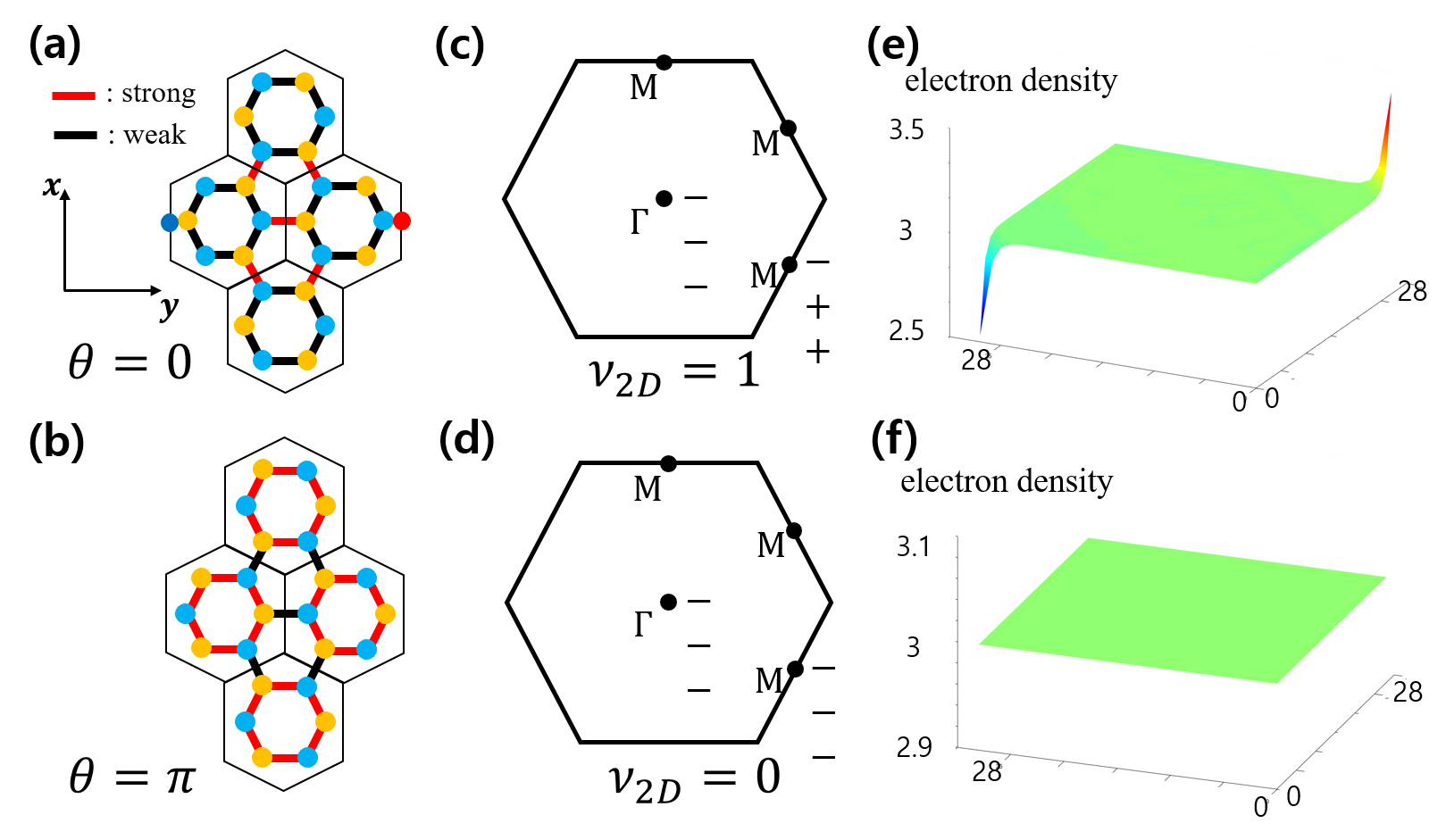}
\caption{(a), (b) Finite-size structure of uniformly textured graphene with $2\times 2$ unit cells. Here blue and yellow dots denote two sublattices. When $\theta=0$, the intercell hopping is bigger than the intracell hopping, and vice versa when $\theta=\pi$. (c), (d) Parity distribution of occupied states at TRIMs for $\theta=0$ (c) and $\theta=\pi$ (d). The parity distributions at three $M$ points are identical.
(e), (f) Charge distribution in a finite-size KTG with $28\times28$ unit cells.
Corner states appear (do not appear) when $\nu_{\rm 2D}=1$ ($\nu_{\rm 2D}=0$).
} \label{fig:topology}
\end{figure}

\section{Higher-order band topology of KTG}
The Hamiltonian for KTG proposed in Ref.~\cite{hou2007electron} is given by
\begin{align}
H= \sum_{\textbf{r}\in \Lambda_{A}}\sum_{j=1}^{3}\left(t+\delta t_j(\textbf{r})\right)c_A^{\dagger}(\textbf{r})c_B(\textbf{r}+\textbf{d}_j)+\textrm{h.c.},
\end{align}
where $c_{A,B}(\textbf{r})$ indicates the (spinless) electron annihilation operator on the sublattice $A$, $B$ at the position $\textbf{r}$, $\textbf{d}_{1,2,3}$ denotes the vectors connecting an $A$ sublattice site to its neighboring $B$ sublattice sites, and $\Lambda_{A}$ denotes the set of the sites belonging to the A sublattice. $t$ denotes the hopping amplitude between nearest neighbor sites, and $\delta t$ describes the modulation of the hopping amplitudes. When $\delta t=0$, the Hamiltonian describes the low-energy band structure of graphene having two Dirac points at the Brillouin zone corners with the momentum $\bm{k}=\pm\bm{K}_{+}$, while $\delta t_j(\textbf{r})=\Delta(\textbf{r}) e^{i\textbf{K}_{+}\cdot\textbf{d}_j}e^{i\textbf{G}\cdot\textbf{r}} + \mathrm{c.c}$ couples the two Dirac fermions at the opposite valleys, where $\textbf{G}=2\textbf{K}_{+}$~\cite{hou2007electron}.
When $\Delta(\textbf{r})$ is a nonzero constant, the unit cell becomes three times larger than that of pristine graphene, and KTG is formed.

Let us examine the band topology of the gapped graphene with a uniform Kekule order parameter. To this end, we define a parametrized Hamiltonian $H_\theta$ describing graphene with a uniform Kekule texture $\Delta(\textbf{r})=\Delta_1+i\Delta_2=\Delta_{0}e^{i\theta}$, where $\Delta_{i=0,1,2}$ and $\theta$ are real constant. As shown in Figs.~\ref{fig:topology}(a) and \ref{fig:topology}(b), we use the convention that the interunit-cell hopping is larger (smaller) than the intraunit-cell hopping when $\theta=0$ ($\theta=\pi$).
When $\theta=0$ or $\pi$, the system is invariant under inversion $P$, the sixfold rotation about the $z$ axis $C_{6z}$, two mirrors $M_x$ and $M_y$ [$M_x:(x,y)\rightarrow (-x,y)$, $M_y:(x,y)\rightarrow (x,-y)$], and time reversal $T$ symmetries. On the other hand, when $\theta\neq 0,\pi$, inversion symmetry is broken, and $H_\theta$ and $H_{-\theta}$
are related by inversion $P$. Also, the chiral (or sublattice) symmetry exists when only the hopping between nearest-neighbor sites is considered.

In general, a 2D $P$-symmetric spinless fermion system carries three $Z_2$ topological invariants, $P_{1x}$, $P_{1y}$, and $\nu_{\rm 2D}$~\cite{hughes2011inversion}.  
Here $P_{1a}$ ($a=x,~y$) indicates the quantized charge polarization along the $a$ direction~\cite{zak1989berry,xiao2010berry},
while $\nu_{\rm 2D}$ is a 2D topological invariant characterizing higher-order topological insulator (HOTI) of $P$-symmetric fermion systems~\cite{ahn2018band, ahn2018higher,ahn2018failure,wieder2018axion}. 
$\nu_{\rm 2D}$ can be determined by using the parity eigenvalues at time-reversal invariant momenta (TRIM) from the following relation
\begin{align}
    (-1)^{\nu_{\rm 2D}}=\prod_{i=1}^{4}(-1)^{[N_{\text{occ}}^{-}(\Gamma_i)/2]},
\end{align}
where $N_{\text{occ}}^{-}(\Gamma_i)$ is the number of occupied bands with odd parity at the TRIM $\Gamma_i$ and the square bracket $[\alpha]$ indicates the integer part of $\alpha$~\cite{song2018diagnosis,fu2007topological,kim2015dirac,hughes2011inversion,turner2012quantized,po2017symmetry}; see Figs.~\ref{fig:topology}(c) and \ref{fig:topology}(d). 

Interestingly, we find that KTG with $\theta=0$ ($\theta=\pi$) is a 2D inversion symmetric HOTI (a trivial insulator) with $\nu_{\rm 2D}=1$ ($\nu_{\rm 2D}=0$)~\cite{noh2018topological, benalcazar2018quantization, geier2018second, khalaf2018higher, ahn2018band}, which is also confirmed by the Wilson loop spectra shown in the Supplemental Material (SM)~\cite{SM, benalcazar2017electric, xie2018second}.
As shown in Fig.~\ref{fig:topology}(e), KTG with $\theta=0$ exhibits a pair of zero-energy corner states related by $P$ symmetry manifesting its higher-order band topology.

\section{Kekule-textured vortex and an axion insulator}
A vortex structure of the order parameter $\Delta(\textbf{r})$ with the winding number $n$ can be introduced by taking $\Delta(\textbf{r})=|\Delta(\textbf{r})|e^{in\theta(\textbf{r})}$, where the polar angle $\theta(\textbf{r})$ varies from $0$ to $2\pi$ encircling the vortex core. 
In particular, when $n=1$, one zero-mode wave function $\psi \sim e^{-\int_{0}^{r}d{r'}|\Delta(r')|}$ is found, yielding a half integral charge $e/2$ localized at the vortex core~\cite{hou2007electron, chamon2008electron, weinberg1981index}.

\begin{figure}[t]
\includegraphics[width=8.5 cm]{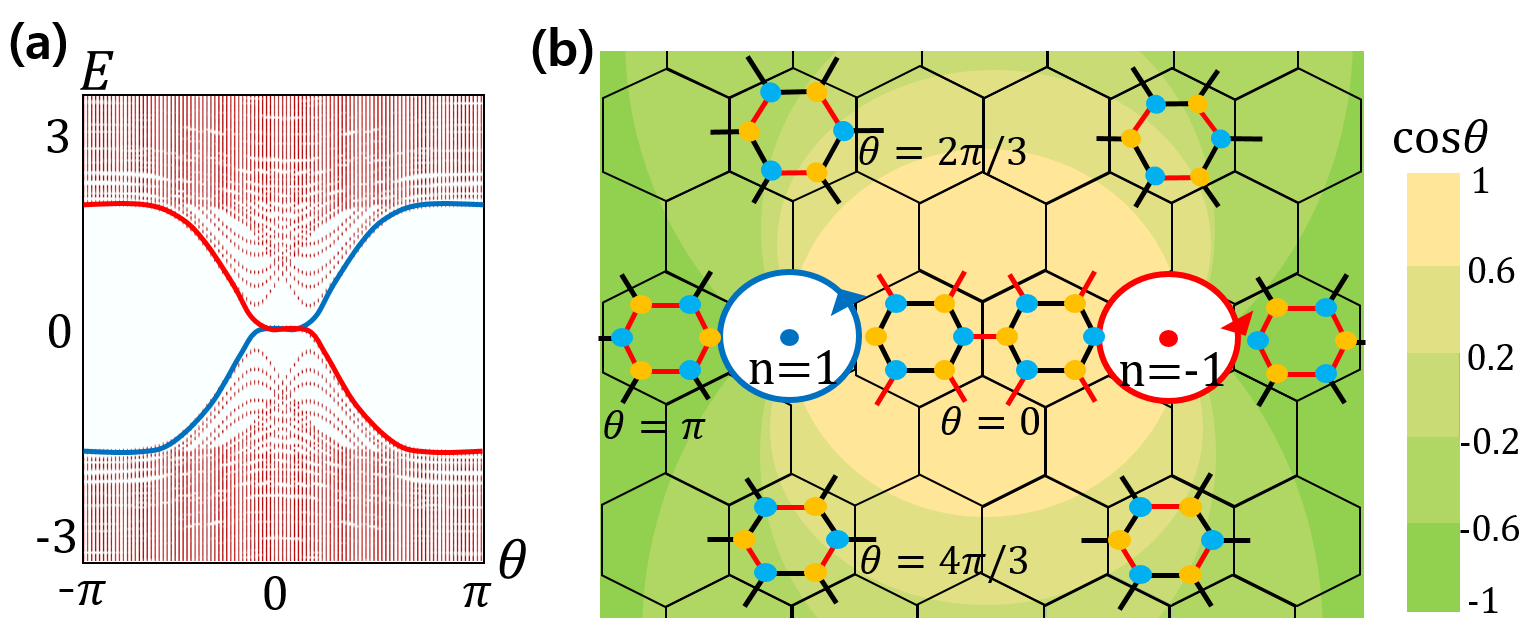}
\caption{(a) Evolution of the energy spectrum of $H_\theta$, with open boundaries in the $x$ and $y$ directions, composed of $28\times 28$ unit cells. 
Two zero-energy modes at $\theta=0$ correspond to the corner charges shown in Fig.~\ref{fig:topology}(e).
During one cycle, one state travels from the valence (conduction) bands to the conduction (valence) bands crossing the Fermi level at $\theta=0$. 
(b) Schematic figure describing the lattice structure around a vortex-antivortex pair in KTG. 
The blue and red circles encircling the vortex and the antivortex, respectively, correspond to two chiral modes (the blue and red lines) in (a). 
Far away from the vortex-antivortex pair, the lattice has a uniform texture with $\theta=\pi$.
} \label{fig:corner}
\end{figure}

The topological property of a vortex can be understood in terms of the charge pumping process of $H_\theta$ during the adiabatic variation of $\theta$ between $0$ and $2\pi$, which corresponds to the phase change $\theta(\textbf{r})$ of the order parameter around a vortex. 
To observe the corresponding spectral flow, we compute the energy spectrum of the Hamiltonian $H_\theta$ under the open boundary condition along the $x$ and $y$ directions [as in the lattices shown in Figs.~\ref{fig:topology}(a) and \ref{fig:topology}(b)]. 
As shown in Fig.~\ref{fig:corner}(a), one eigenstate travels from the valence (conduction) bands to the conduction (valence) bands during the variation of $\theta$ from $-\pi$ to $\pi$. 
Two zero-energy modes at $\theta=0$ account for the corner charges shown in Fig.~\ref{fig:topology}(e).

The topological origin of the nontrivial spectral flow can be understood from the symmetry of $H_\theta=\sum_{\textbf{k}}H(\textbf{k};\theta)$ as follows. Here $\textbf{k}=(k_x,k_y)$ denotes a momentum in the 2D Brillouin zone.
Under inversion $P$, the parametrized Hamiltonian $H(k_x,k_y;\theta)$ transforms as
\begin{align}
    PH(k_x,k_y;\theta)P^{-1}= H(-k_x,-k_y;-\theta).
\end{align}
If $\theta$ is taken as a third momentum $k_z$, the effective 3D Hamiltonian $H_{\text{3D}}(\bm{k})\equiv H(k_x,k_y;\theta=k_z)$ can be considered as a Hamiltonian for a $P$ invariant 3D insulator.
Let us note that $H_{\text{3D}}(\bm{k})$ restricted in the 2D $P$-invariant momentum subspaces with $k_z=0$ and $\pi$ is characterized by the 2D $\mathbb{Z}_2$ invariant $\nu_{\rm 2D}(k_z)$, as the 2D Hamiltonian $H(k_x,k_y;\theta)$ with $\theta=0$ and $\theta=\pi$ are the $P$ invariant models with a Kekule texture.
For a 3D insulator with $P$ symmetry, it was recently shown \cite{ahn2018higher,wieder2018axion} that the quantized magnetoelectric polarizability $P_3$ is equivalent to $\Delta \nu_{\rm 2D}=\nu_{\rm 2D}(k_z=\pi)-\nu_{\rm 2D}(k_z=0)$, namely,
\begin{align}
\Delta \nu_{\rm 2D} =2P_3= \frac{1}{4\pi^2}\int_{T^2\times S^{1}}\text{Tr}\!\left[AdA-\frac{2i}{3}A^3\right] \mod 2,
\end{align}
where $A_{ij}=i\langle u_i|du_j\rangle$ is the non-Abelian Berry's connection characterizing the valence band eigenstates $|u_i\rangle$, $T^{2}$ indicates the 2D Brillouin zone, $S^1$ denotes the unit circle parametrized by $\theta$.
Therefore, when $\Delta \nu_{\rm 2D}=1$ (mod 2), the Hamiltonian $H_{\text{3D}}(\bm{k})$ exhibits quantized $P_3=1/2$, that is, $H_\mathrm{3D}(\bm{k})$ describes an axion insulator~\cite{qi2008topological,turner2012quantized}. 
Due to the nontrivial band topology of $H_{\text{3D}}(\bm{k})$, the energy spectrum of the Hamiltonian under open boundary conditions in the $x$ and $y$ directions exhibits a topological spectral flow as a function of $k_z$. Namely, one electron should be pumped from the valence bands to the conduction bands and vice versa during the variation of $k_z\in[-\pi,\pi]$, which corresponds to two chiral hinge modes of the axion insulator described by $H_\mathrm{3D}(\bm{k})$.

The topological spectral flow of $H_\mathrm{3D}(\bm{k})$ explains the charge accumulation at a vortex as follows. Let us consider a textured lattice with a vortex-antivortex pair shown in Fig.~\ref{fig:corner}(b), where the parameter $\theta$ is fixed to be $\theta=\pi$ far away from the vortices. 
Since $H_{\theta}=\pi$ is topologically trivial, the region with $\theta=\pi$ can be considered as a vacuum. 
Then, in the region with $\theta=0$ between the vortex and the antivortex, the centers of the vortex and the antivortex can be considered as two corners of $H_{\theta=0}$ under open boundary condition; See Fig.~\ref{fig:topology}(a). Therefore, the vortex bound state can be considered as the corner state of the HOTI.

The spectral flow in Fig.~\ref{fig:corner}(a) implies that an in-gap state travels from the valence (conduction) to the conduction (valence) bands when $\theta$ changes from $0$ to $2\pi$ along a circle enclosing a vortex (antivortex). This means that the in-gap state should cross the Fermi level in the middle of the variation of $\theta\in[-\pi, \pi]$, leading to a zero-energy bound state at the core of the vortex (antivortex). This shows that the topological charge pumping process underlies the relationship between the existence of in-gap states and the bulk band topology of KTG. 
We provide a field theoretical explanation for the connection between topological spectral flow and charge accumulation at a vortex in SM~\cite{SM}. 

\section{Generalization}
A similar mechanism for charge accumulation at a vortex in a textured lattice due to the topological spectral flow can be applied to any parametrized 2D Hamiltonian $H(k_x,k_y;\theta)$ when its corresponding 3D Hamiltonian $H_{\text{3D}}(\bm{k})$ exhibits quantized $P_3$. 
In the following discussion, the presence or absence of spin-orbit coupling makes no difference unless noted otherwise.
Let us note that $P_3$ is quantized when the 3D insulator described by $H_{\text{3D}}(\bm{k})$ has a space-time orientation reversing symmetry such as $\mathcal{T}$, $\mathcal{P}$, $\mathcal{C}_{n}\mathcal{T}$, and $\mathcal{C}_{n}\mathcal{P}$ ($n=2,3,4,6$)~\cite{varjas2015bulk,qi2008topological,schindler2018higher2,ahn2018higher, ryu2010topological}, where we have used calligraphic fonts for symmetries in the 3D space to distinguish them from the symmetries in the 2D space denoted in italic fonts.
Among these symmetries, we can neglect $\mathcal{C}_{3z}\mathcal{T}$ in that a $\mathcal{C}_{3z}\mathcal{T}$-symmetric axion insulator is expected to carry the same topological properties as the one with $\mathcal{T}$ symmetry only, because $(\mathcal{C}_{3z}\mathcal{T})^3=\mathcal{T}$ and $(\mathcal{C}_{3z}\mathcal{T})^2=\mathcal{C}_{3z}^{-1}$, and $\mathcal{C}_{3z}$ by itself cannot quantize $P_3$. Likewise, we do not have to consider $\mathcal{C}_{6z}\mathcal{T}$ since $(\mathcal{C}_{6z}\mathcal{T})^3=\mathcal{C}_{2z}\mathcal{T}$. For $\mathcal{C}_{nz}\mathcal{P}$ symmetries, $(\mathcal{C}_{3z}\mathcal{P})^3=\mathcal{P}$ and $(\mathcal{C}_{6z}\mathcal{P})^3=\mathcal{C}_{2z}\mathcal{P}=\mathcal{M}_z$. Thus, we only have to consider $H_{\text{3D}}(\bm{k})$ with $\mathcal{T}$, $\mathcal{P}$, $\mathcal{C}_{2z}\mathcal{P}$, $\mathcal{C}_{2z}\mathcal{T}$, $\mathcal{C}_{4z}\mathcal{P}$, or $\mathcal{C}_{4z}\mathcal{T}$ symmetries.
It is worth noting that in each axion insulator described by $H_{\text{3D}}(\bm{k})$, the 2D subspaces with $k_z=0$ and $k_z=\pi$, respectively, support distinct 2D topological invariants. Namely, the quantized $P_3$ of $H_{\text{3D}}(\bm{k})$ can be obtained from the difference between the 2D topological invariants on the two symmetry invariant planes with $k_z=0$ and $k_z=\pi$, respectively~\cite{fu2006time, bernevig2006quantum, teo2008surface, van2018higher, schindler2018higher}. This ensures that the corresponding vortex structure described by $H(k_x,k_y,\theta)$, which connects two distinct topological phases at $\theta=0$ and $\pi$, hosts a fractional vortex charge.

\begin{table}[t]
\caption{
The correspondence between the symmetry of $H_{\text{3D}}(\bm{k})$ exhibiting quantized $P_3$ and that of the 2D Hamiltonian $H(k_x,k_y;\theta=0,\pi)$.  For $T$ and $C_4T$ symmetry, spin orbit coupling (SOC) is necessary for having $P_3=1/2$, while SOC is not required for other symmetries. (See the SM). The nature of the corresponding 2D topological insulator (TI) is shown in the fourth column. Here ``QSHI" indicates a quantum spin Hall insulator.}
\begin{tabular}{c | c | c | c}
\hline
\hline
3D Symmetry & 2D Symmetry & SOC & 2D TI\\
\hline
\hline
$\mathcal{P}=\mathcal{C}_{2z}\mathcal{M}_z$ & $P$, $C_{2z}$, $C_{6z}$ & & HOTI
\\
$\mathcal{C}_{4z}\mathcal{P}=(\mathcal{C}_{4z}\mathcal{M}_z)^{3}$  &$C_{4z} $ & & HOTI
\\
$\mathcal{T}$ & $T$ & \checkmark & QSHI
\\
$\mathcal{M}_z$ & $M_z$ &  & Mirror TI
\\
$\mathcal{C}_{2z}\mathcal{T}$ & $PT$, $C_{2z}T$, $C_{6z}T$ & & HOTI 
\\
$\mathcal{C}_{4z}\mathcal{T}$ & $C_{4z}T$ & \checkmark & HOTI 
\\
\hline \hline
\end{tabular}
\end{table}\label{table:1}

Now let us explain how the symmetry of the 3D Hamiltonian $H_{\text{3D}}(\bm{k})$ can be related to that of the physical 2D parametrized Hamiltonian $H(k_x,k_y;\theta)$. For instance, as in the case of KTG, if the system with a uniform order parameter is invariant under $n$-fold rotation $C_{n}$ about the $z$ axis at $\theta=0$ and $\pi$, but not at other $\theta$ values, then the $C_{n}$ symmetry can be implemented as
\begin{align}\label{eq:Cn}
C_nH(k_x,k_y;\theta)C_n^{-1}=H(k'_x,k'_y;-\theta),
\end{align}
where $k'_x$, $k'_y$ are the rotated momenta after $C_{n}$ operation.
Due to the sign change of $\theta$ under $C_n$, the 2D system is $C_n$ invariant only at $\theta=0,~\pi$. Then the $C_n$ symmetry of the 2D system can be implemented in $H_{\text{3D}}(\bm{k})$ as the $\mathcal{C}_n\mathcal{M}_z$ symmetry. Now we ask whether $\mathcal{C}_n\mathcal{M}_z$ symmetry can quantize $P_3$ and also whether $C_{n}$ symmetry supports a 2D topological invariant $\nu_{\rm 2D}$ such that its difference $\Delta\nu_{\rm 2D}=\nu_{\rm 2D}(\theta=\pi)-\nu_{\rm 2D}(\theta=0)$ is identical to $2P_3$ (modulo two).
Similar ideas can be applied to all the other symmetries that can give quantized $P_3$~\cite{SM}.



Table I summarizes the correspondence between the symmetry of the 2D insulator with a uniform gap-opening order parameter like the Kekule texture, described by the Hamiltonian $H(k_x,k_y;\theta=0,\pi)$, and the symmetry of the relevant 3D Hamiltonian $H_{\text{3D}}(\bm{k})$ exhibiting quantized $P_{3}$.
When the two Hamiltonian $H(k_x,k_y;\theta=0)$ and $H(k_x,k_y;\theta=\pi)$ support distinct bulk topological properties, a fractional charge can be localized at a vortex of the order parameter parametrized by $\theta$ due to the topological spectral flow associated with the quantized $P_{3}$. 

\section{Chiral symmetry and charge quantization}
In fact, the electric charge accumulated at a vortex is half-quantized only when chiral symmetry exists. When chiral symmetry is broken, the midgap states can be shifted from the zero energy and the vortex charge takes an arbitrary value. 
Yet, the topological invariant $P_3=\nu_{\rm 2D} (\theta=\pi)-\nu_{\rm 2D} (\theta=0)$ is quantized irrespective of the presence or absence of chiral symmetry, and this guarantees a quantized half-integral \textit{Wannier} charge localized at a vortex as long as the relevant crystalline symmetry is preserved.

\begin{figure}[t]
\includegraphics[width=8.5 cm]{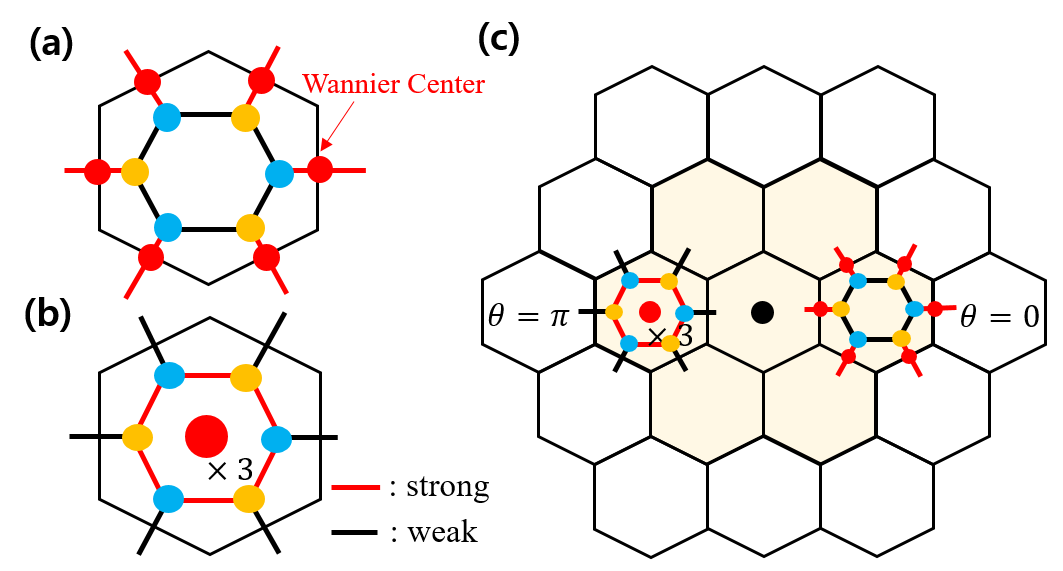}
\caption{(a) The location of Wannier centers when $\theta=0$ corresponding to the 2D HOTI with $\nu_{\rm 2D}=1$. (b) A similar figure when $\theta=\pi$ corresponding to a trivial insulator with $\nu_{\rm 2D}=0$.
(c) Distribution of Wannier centers around a vortex. The Wannier functions centered at the unit cell boundary when $\theta=0$ contributes the quantized half-integer Wannier charge in the shaded region including the vortex core.  
} \label{fig:Wannier}
\end{figure}

To define a Wannier charge, we assume that an electron is sharply localized at the Wannier center for the counting of electron numbers, and the Wannier centers are located at the Wyckoff positions that are invariant under the crystalline symmetres up to a lattice constant vector. Then the electron number sharply localized at a Wannier center is the Wannier charge.


The quantized Wannier charge can be seen from the distribution of Wannier center positions described in Fig.~\ref{fig:Wannier}. When $\theta=0$ ($\theta=\pi$) with $P$ symmetry, the locations of Wannier centers are fixed at the boundary (center) of the unit cell. Thus, these two states cannot be adiabatically connected when $P$ exists. On the other hand, for other $\theta$ with broken $P$, the Wannier centers are located at generic points within a unit cell, and always contribute an integer number of electrons. 
Thus, a group of connected unit cells including the vortex core, such as the shaded region in Fig.~\ref{fig:Wannier} (c), always contains a half-quantized Wannier charge
due to the Wannier centers at the unit cell boundary when $\theta=0$. 
This is similar to the situation of the $P$ symmetric
SSH chain having quantized charge polarization, which gives
a half-integral Wannier charge at the domain wall irrespective
of chiral symmetry~\cite{SM}. 

This Wannier charge description can generally be applied to the cases with other symmetries~\cite{SM}. 
Let us note that the topological properties of the 2D HOTI lacking chiral symmetry, especially its relation with the third-order TI and the related filling anomly, have been extensively studied in recent papers~\cite{cualuguaru2019higher, okuma2019topological, song2017d, lee2020two,hwang2019fragile}.

\section{Discussion}
In real materials, the $U(1)$ symmetry of the order parameter may be reduced to a discrete $\mathbb{Z}_{n=2,3,4,6}$ symmetry by lattice potentials. However, fractional charges can still be localized at a vortex with discrete symmetry as long as the phase $\theta$ of the order parameter winds $2\pi$ around the vortex. For instance, in the case of KTG, fractional charge can be localized at a junction where three domains with $\theta=0$, $\pi/3$, $5\pi/3$, respectively, meet. Alternatively, one can construct heterostructures composed of 2D topological and normal insulators to observe fractional bound charges~\cite{SM, qi2008fractional}.

One interesting direction for future research is to extend the idea of symmetry protected topological vortices to defects of various codimensions. For example, it is shown in Ref.~\cite{teo2010topological} that a point defect carrying fractional charges in 3D systems can be described by using the Hamiltonian for five-dimensional insulators whose topological invariant has the Chern-Simons $5$-form. Systematic classification of such defect structures taking into account space group symmetries is desirable to complete the classification table for defect Hamiltonian beyond the tenfold classification scheme proposed before Ref.~\cite{teo2010topological}.

\begin{acknowledgments}
E.L. was supported by IBS-R009-D1.
A.F. was supported by JSPS KAKENHI (Grant No.\ 15K05141 and 19K03680) and JST CREST (Grant No.~\mbox{JPMJCR19T2}).
B.-J.Y. was supported by the Institute for Basic Science in Korea (Grant No.~IBS-R009-D1) and Basic Science Research Program through the National Research Foundation of Korea (NRF) (Grant No.~0426-20180011), and  the POSCO Science Fellowship of POSCO TJ Park Foundation (No.~0426-20180002).
This work was supported in part by the U.S. Army Research Office under Grant No. W911NF-18-1-0137. 
We appreciate the helpful discussions with J. Ahn and Y. Hwang.
\end{acknowledgments}

\setcounter{section}{0}
\setcounter{figure}{0}
\setcounter{equation}{0}

\renewcommand{\thefigure}{S\arabic{figure}}
\renewcommand{\theequation}{S\arabic{equation}}
\renewcommand{\thesection}{S\arabic{section}}

\tableofcontents
\section{Two different ways of describing charge fractionalization in SSH model}
Here let us review the physics of charge fractionalization described by the Su-Schriefer-Heeger (SSH) model~\cite{su1980soliton}. The Hamiltonian can be written as
\begin{align}
    H_{\textrm{SSH}}=\sum (t+\delta t_i)a^{\dagger}_{i}a_{i+1} +H.c,
\end{align}
where $t$ is the hopping amplitude and its modulation is described by $\delta t_i = (-1)^{i}\Delta$. Also, we assume that the system is half-filled.
If $\Delta$ is 0, the energy dispersion is gapless and the unit cell contains one atom. However, when $\Delta$ is non-zero constant, the unit cell becomes doubled so that the size of the Brillouin zone is reduced by half and gapped band structure appears. Spontaneous formation of lattice dimerization gives rise to two degenerate ground states distinguished by the opposite sign of the mass term $\Delta$.
Depending on the sign of $\Delta$, the system becomes either trivial or topological. In our convention, $\Delta>0$ corresponds to the topological phase and $\Delta<0$ corresponds to the trivial phase.
At the boundary between topological and trivial phases, there exists a localized zero-mode wave function and the bound charge is quantized to 1/2. Let us introduce two different ways describing charge fractionalization at the domain wall.

In the first approach, we solve the low energy Dirac Hamiltonian given by
\begin{equation}
    H(x)= 
\left(\begin{array}{cc} -i\partial_x & \Delta(x)\\ \Delta^*(x) & i\partial_x \end{array}\right).
\end{equation}
Across the domain wall, $\Delta(x)$ changes from $-\Delta_0$ to $\Delta_0$, and it is complex in between.
The equation satisfied by the zero-mode solution is $H(x)\psi(x)=0$ where $\psi(x)=(u(x), v(x))^{T}$. More explicitly,
\begin{align}
    -i\partial_x u(x) + \Delta(x) v(x) = 0, \nonumber \\ 
    \Delta^*(x) u(x) +i\partial_x v(x) = 0. 
\end{align}
Under the solitonic mass background, one can find a normalizable zero energy wave function $\psi(x)$ $\sim e^{-\int_0^x |\Delta(x')|dx'}$, which gives a half electric charge localized at the domain wall.

In the second approach, one can explain the existence of fractional charge in terms of the bulk topological property. Let us consider a Hamiltonian $H(k_x, \theta)=\sin{k_x}\sigma_z+\Delta_x(k_x,\theta)\cos\theta\sigma_x+\Delta_y(k_x,\theta)\sin\theta\sigma_y$ that varies adiabatically along a circle parametrized by $\theta\in[0,2\pi]$ without closing a bulk gap, as in Ref.~\onlinecite{teo2010topological} (See Fig. S1(a)). We assume that Hamiltonian describes a trivial (topological) phase when $\theta=0$ $(\pi)$ with the corresponding polarization $P_1=0$ ($P_1=1/2$), where $P_1=\frac{1}{2\pi}\int_{T^1} \text{Tr}[A]$ is quantized to be either $0$ or $1/2$ $\mod 1$ due to the inversion $P$ symmetry. 
On the other hand, $P_1$ is not quantized for the states with $\theta\neq0,~\pi$, since $P$ is broken due to the term proportional to $\sigma_y$.

In the presence of $P$ symmetry, $P_1$ can also be computed by multiplying the parity eigenvalues at the time reversal invariant momenta (TRIM)~\cite{hughes2011inversion, turner2012quantized} as
\begin{align}
    (-1)^{2P_1}=(-1)^{N_{\text{occ}}^{-}(k=0)}(-1)^{N_{\text{occ}}^{-}(k=\pi)},
\end{align}
where $N_{\text{occ}}^{-}(\Gamma_i)$ is the number of occupied states with negative parity at $\Gamma_i$.
\begin{figure}[t]
\centering
\includegraphics[width=8.5 cm]{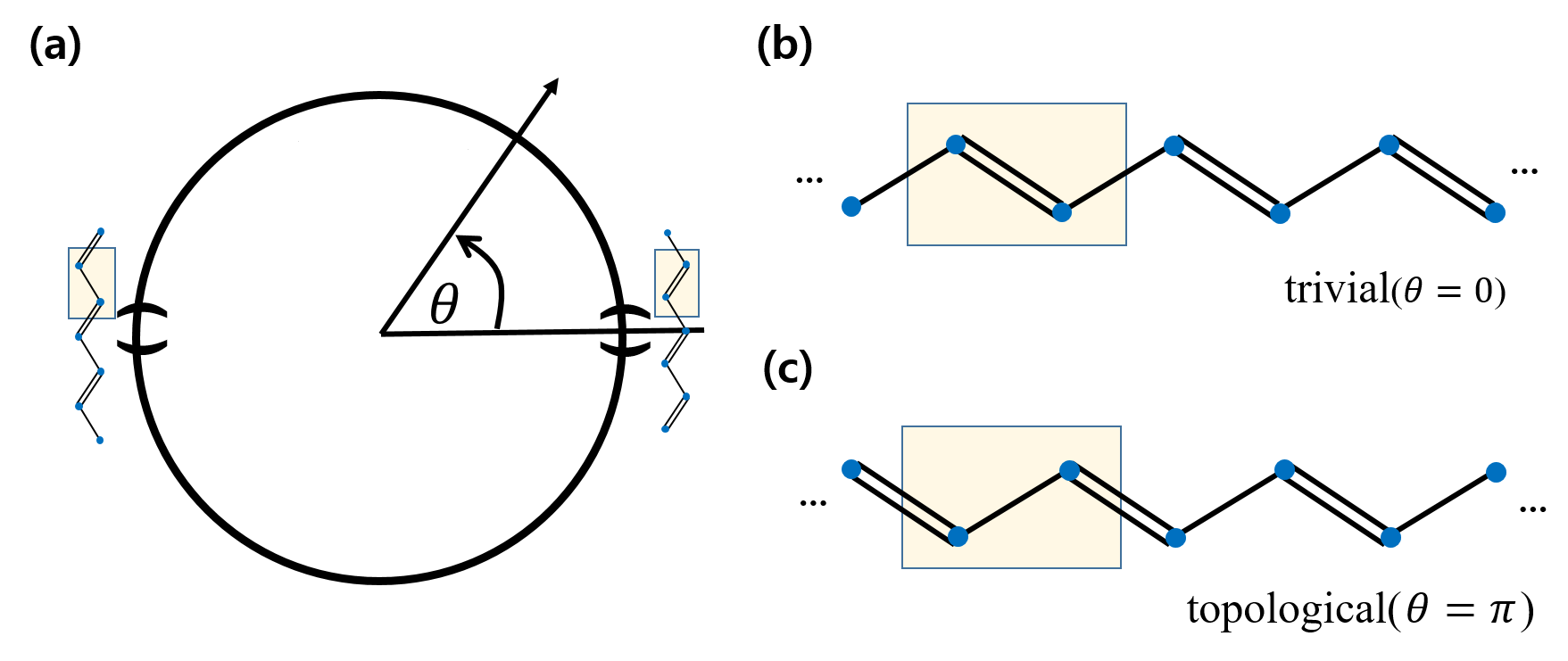}
\caption{(Color online) (a) An adiabatic cycle interpolating two insulators with the charge polarization $P_1=0$ and $P_1=1/2$, respectively. (b) Schematic figures describing the trivial (topological) phase with the parameter $\theta=0$ ($\theta=\pi$)
} \label{fig:SSH}
\end{figure}
The bound charge between two insulators with $P_1=0$ ($\theta=0$) and $P_1=1/2$ ($\theta=\pi$) is expressed by the integral of the polarization gradient
\begin{align}
    q_{\text{bound}}=\int -\partial_\theta P_1 d\theta = 1/2\textrm{ mod 1},
\end{align}
which indicates the half integral electric charge localized between two phases.

While the parametrized Hamiltonian $H(k_x,\theta)$ is $P$ invariant only at $\theta=0$ and $\pi$, $P$ symmetry can be implemented as
\begin{align}
    PH(k_x,\theta)P^{-1}= H(-k_x,-\theta).
\end{align}
Interestingly, when $\theta$ is replaced by $k_y$, the Hamiltonian $H(k_x,k_y)$ describes a 2D Chern insulator with $P_{2D}$ symmetry,
\begin{align}
    P_{2D}H(k_x,k_y)P_{2D}^{-1}= H(-k_x,-k_y).
\end{align}
This is because the Chern number C can be related to the change of the charge polarization through the parity formula for occupied bands as
\begin{align}
(-1)^{C}=(-1)^{\sum N_{\text{occ}}^{-}(\Gamma_i)}=(-1)^{2P_1(\theta=\pi)-2P_1(\theta=0)},
\end{align}
where $\Delta (2P_1)=C \mod 2$. Thus, $\Delta (2P_1)=1$ implies that there is a nontrivial spectral flow from the conduction band to the valence band during the adiabatic evolution of $\theta$ from $0$ to $2\pi$, which corresponds to the chiral edge mode of the Chern insulator~\cite{qi2008topological}.
Since the chiral edge mode connects the valence and conduction bands, even if chiral symmetry is broken, the half charge should be localized at the domain wall between the topological and trivial phases in $P$ symmetric SSH model.

As shown in the main text, the fractional charges bound to the vortex of the order parameter can also be explained by the bulk topological response of the effective Hamiltonian for axion insulators in three dimensions. Table I compares the charge fractionalization at the domain wall of the SSH model and at the order parameter vortex of the HCM model.

\begin{table}[h]
\begin{tabular}{ c ||c| c }
\hline
\hline
& Domain wall of SSH model & Vortex of HCM model \\
\hline
\hline
$H_{\rm D+1}$ & Chern insulator  & Axion insulator  
\\
Invariant & $C_1=\Delta (2P_1)$  & $2P_3=\Delta \nu_{2D}$   
\\
\hline \hline
\end{tabular}
\caption{Comparison of the charge fractionalization mechanism at a domain wall in 1D systems described by the SSH model and that at an order parameter vortex in 2D systems described by the HCM model. Fractional charges localized at a point defect in $D$-dimensional systems can be explained by the topological response of the $D+1$ dimensional effective Hamiltonian $H_{D+1}$.}
\end{table}\label{table:2}

\section{Nested Wilson loop}
A Wilson loop is a gauge invariant observable whose eigenvalue spectrum contains the information on the topological properties of the Hamiltonian. 
In $2D$, a Wilson loop operator is defined by
\begin{align}
W_{(k_1+2\pi,k_2)\leftarrow(k_1,k_2)}&\equiv W_{x,\textbf{k}}
\nonumber\\
&= \lim_{N\rightarrow\infty}F_{N-1}F_{N-2}\dotsb F_{1}F_{0}
\nonumber\\
&=Pe^{-i\oint_CA_kdk},
\end{align}
where $[F_i]_{nm}=\langle u_m(k_{i+1},k_2)|u_n(k_{i},k_2)\rangle, k_i=\frac{2\pi}{N}i,$ and $m,n=1,\dotsb,N_{\rm occ},  i=1,\dotsb,N$.
Since a Wilson loop operator is unitary, the eigenvalue equation is given by $W_{x,\textbf{k}}|\nu_{x,\textbf{k}}^{\ell}\rangle=e^{i\nu_x^{\ell}(k_y)}|\nu_{x,\textbf{k}}^{\ell}\rangle$, where $\nu_x^{\ell}(k_y)$ corresponds to the $x$-component of the Wannier center of $\ell$th Wannier functions.
It follows that the electron charge polarization is expressed as 
\begin{align}
p_x&=\frac{1}{2\pi N_y}\sum_{k_y}\sum_{\ell=1}^{N_{occ}}\nu_{x}^{\ell}(k_y)\nonumber\\&=-\frac{i}{2\pi N_y}\sum_{k_y}\log\det[W_{x,\textbf{k}}]=-\frac{1}{2\pi}\oint \text{Tr}[A_{x,k}]d^2k. 
\end{align}
In the presence of $P$ symmetry, the set of eigenvalues satisfy $\{\nu_x(k_y)\}\equiv\{-\nu_x(-k_y)\}$ mod $2\pi$, so that the polarization is quantized to either $0$ or $1/2$ modulo 1.

Recently, the nested Wilson loop method was developed to study higher order topological properties~\cite{benalcazar2017quantized, benalcazar2017electric}. The procedure for computing the nested Wilson loop is as follows.
First, we calculate a Wilson loop operator along a reciprocal vector $\bf{G_1}$, where its eigenvalues are $\{e^{i\nu_{1}(k_2)}\}$. Next, we choose a certain subset of Wilson loop eigenvalues $\{\nu_{1}\}$ and find the corresponding Wannier eigenfunctions$\{|\nu_{1}(k_2)\rangle\}$. With the eigenfunctions, we calculate a Wilson loop operator along the other reciprocal vector $\bf{G_2}$, where we obtain a nested Wilson loop operator $\widetilde{W}$ as
\begin{align}
\widetilde{W}_{k_2+2\pi\leftarrow k_2}=\widetilde{W}_2=\lim_{N\rightarrow\infty}\widetilde{F}_{N-1}\widetilde{F}_{N-2}\dotsb \widetilde{F}_{1}\widetilde{F}_{0},
\end{align}
where $[\widetilde{F}_i]_{nm}=\langle \nu_m(k_{i+1})|\nu_n(k_{i})\rangle, k_i=\frac{2\pi}{N}i$ and $m,n=1,\dotsb,N_{\text{sub}}$. Here $N_{\text{sub}}$ denotes the number of the subbands of a gapped Wilson loop spectrum.
By using this method, it is possible to detect the electric multipole moments of the system~\cite{benalcazar2017quantized,benalcazar2017electric,xie2018second}.

\begin{figure}[h]
\centering
\includegraphics[width=8.5 cm]{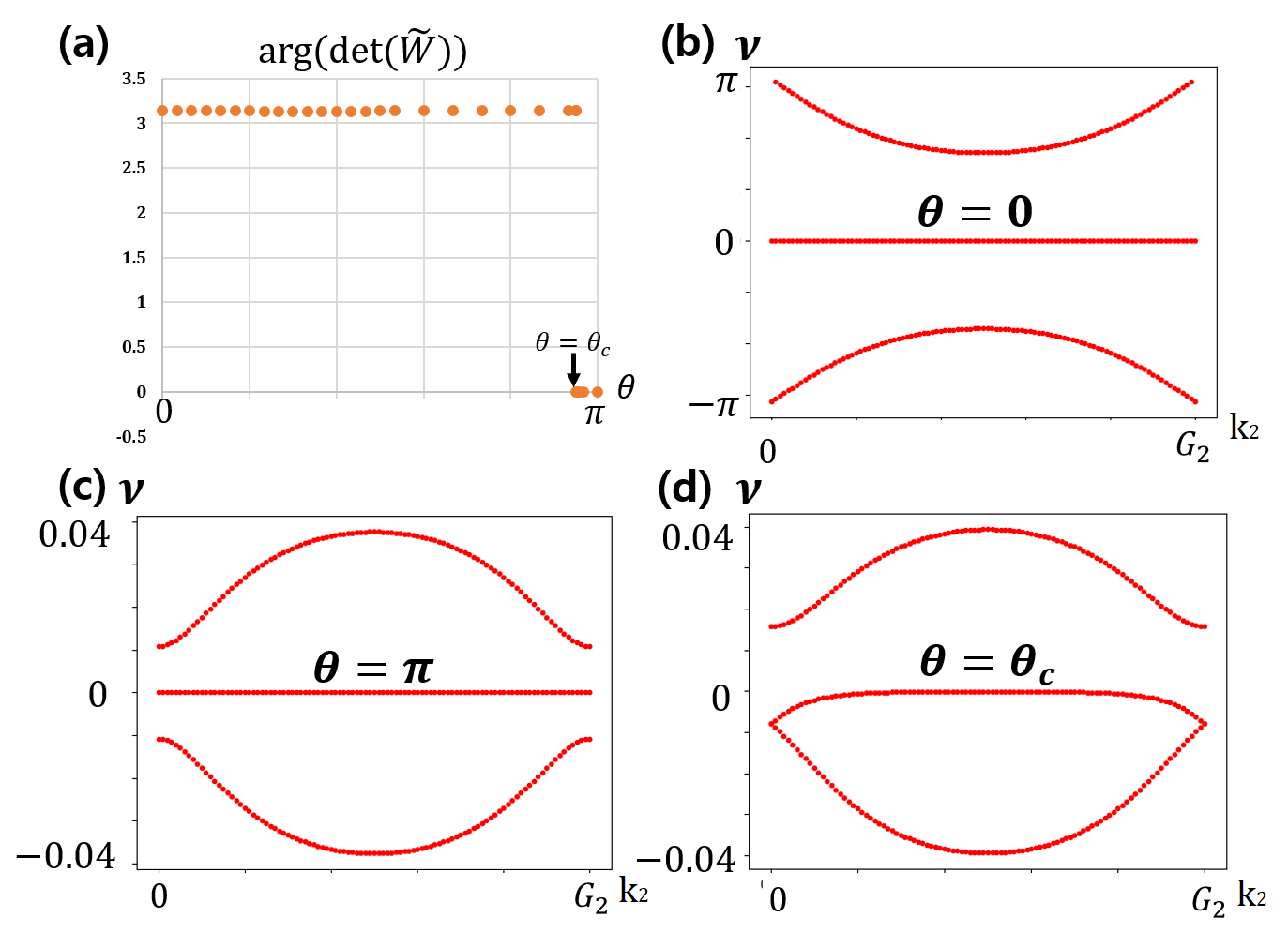}
\caption{(Color online) (a) The determinant of nested Wilson loops for the model Hamiltonian describing the Kekule textured graphene, which is plotted for $0\leq\theta\leq\pi$. At $\theta=\theta_c$, $\det(\widetilde{W})$ changes abruptly. (b) The Wilson loop spectrum for $\theta=0$. The spectrum has a Dirac-like crossing point at $k=0$, and the determinant of the nested Wilson loop is $-1$.  (c) The Wilson loop spectrum for $\theta=\pi$. The Wilson loop spectrum of a trivial insulator has no band crossing and the determinant of the nested Wilson loop is $1$. (d) The Wilson loop spectrum for $\theta=\theta_c$, where the gap is closed, which implies the phase transition between $\det(\widetilde{W})=-1$ and $+1$.
} \label{fig:wilson}
\end{figure}
The Kekule textured graphane has inversion $P$ symmetry, where the Wilson loop spectrum satisfies $\{\nu_{1}(k_2)\} \equiv \{-\nu_{1}(-k_2)\} \mod 2\pi$. Also, it has spinless $T$ symmetry, where the Wilson loop spectrum satisfies $\{\nu_{1}(k_2)\} \equiv \{\nu_{1}(-k_2)\} \mod 2\pi$. 
Put together,  $\{\nu_{1}(k_2)\} \equiv \{-\nu_{1}(k_2)\} \mod 2\pi$.
Let us note that $T$ is not crucial in quantizing $\det{\widetilde{W}}$, but we have introduced the symmetry for the sake of simplicity.
Thus, a Wilson band must be $\nu_{1}(k_2)=0$, $\pi$ or exist as a pair $\{+\nu_1(k_2), -\nu_1(k_2)\}$, so that the spectrum can be divided into two subsets that are centered at either $\nu_1=0$ or $\nu_1=\pi$~\cite{ahn2018failure}.
We choose two symmetric Wannier bands that are above and below $\nu_1=\pi$, and calculate the relevant nested Wilson loop $\widetilde{W}$. 
The associated Wannier sector polarization is $p_{y}^{\nu_1}=-\frac{i}{2\pi}\log\det[\widetilde{W}]$.
Under $PT$ symmetry, $p_2^{\nu_1}$ is quantized to $1/2$ or $0$ since $p_2^{\nu_1}\equiv-p_2^{-\nu_1}$ mod 1~\cite{benalcazar2017quantized,benalcazar2017electric}.
Also, it is known that $\det{\widetilde{W}}$ is equal to $(-1)^{\nu_{2D}}$~\cite{ahn2018failure}.
Thus, $\nu_{2D}=2p_2 \mod 1$.

For the Kekule textured graphene, $P$ exists when $\theta=0$ or $\pi$.
In Fig.~\ref{fig:wilson}, the Wilson loop spectrum of Kekule textured graphene is shown, where $\det(\widetilde{W})$ is quantized as $-1$ for $\theta=0$ while $\det(\widetilde{W})$ is quantized as $1$ for $\theta=0$.
As $\theta$ varies, we observe that the Wilson loop spectrum undergoes a phase transition between $\theta=0$ and $\theta=\pi$. $\det(\widetilde{W})$ changes abruptly and the gap of the Wilson spectrum is closed at a critical point.

\section{Influence of core electrons to the higher order band topology}
\begin{figure}[t]
\centering
\includegraphics[width=8.5 cm]{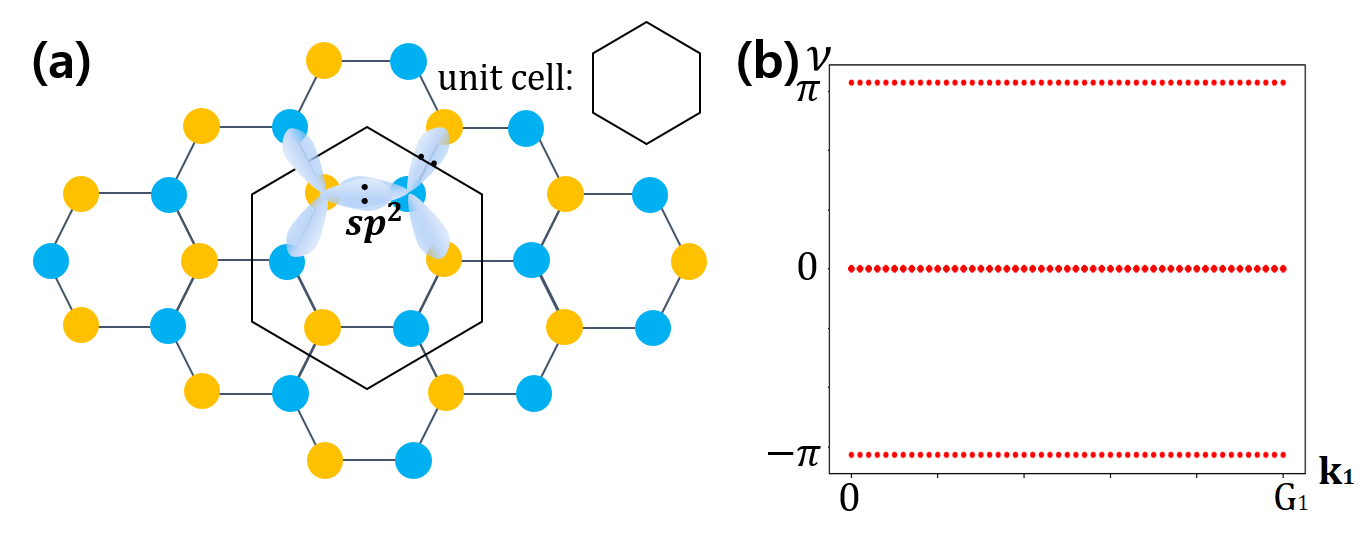}
\caption{(Color online) (a) $sp^2$ orbitals on the honeycomb lattice. When only nearest neighbor hopping is considered, electrons are localized at $\sigma$ bonds. (b) The Wilson loop spectrum for core levels composed of $sp^2$ orbital exhibiting $\nu_{2D}=1$. There are nine Wilson bands. Seven of them are at $\nu = 0$ and two of them are at $\nu=\pi$.
} \label{fig:sp2}
\end{figure}
To construct the tight-binding Hamlitonian for the Kekule textured graphene, only $p_z$ orbitals are taken into account since they are mainly responsible for the low energy electronic property near the Fermi level. Core energy levels that are far away from the Fermi level are assumed to have negligible effect.
However, when the bulk topological invariant is concerned, core levels far below the Fermi level can also play a crucial role. For instance, in graphene, $sp^2$ orbitals form core levels below the Fermi level since the $\sigma$ bonding between $sp^2$ orbitals is stronger than the $\pi$ bonding between $p_z$ orbitals.
To see the effect of core energy level in higher order topological physics, let us construct a simple tight-binding Hamiltonian for $sp^2$ orbitals, which only takes into account nearest neighbor hopping, 
\begin{align}
    H=\sum_{n=1}^{9} t_{\sigma}a_{2i-1}^{\dagger}a_{2i} +H.c,
\end{align}
where $t_{\sigma}$ is hopping parameter for a $\sigma$ bonding and $a_{i}$ is the annihilation operator of the $i$-th $sp^2$ orbital. There are 9 $sp^2$ orbitals filled in the unit cell as shown in Fig.~\ref{fig:sp2}(a).
The energy eigenvalues are degenerate and the electronic bands are dispersionless because electrons are localized between neighboring two atoms.
From the Wilson loop spectrum or the parity eigenvalues at TRIM points, the $2D$ topological invariant $\nu_{2D}$ turns out to be nontrivial.
Thus, in Kekule textured graphene, when the lower energy levels are included, the Kekule texture with $\theta=0$ becomes trivial ($\nu_{2D}=0$) while that with $\theta=\pi$ becomes nontrivial ($\nu_{2D}=1$) (See Fig.~\ref{fig:sp2}(b)). 
However, the change of $\nu_{2D}$ remains the same, and thus fractional charges bound to vortices are also robust against the adding additional bands below the Fermi level.

\section{Second- and third-order topological insulators}
The zero-mode corner states of a higher-order topological insulator (HOTI) in 2D systems are protected only if chiral symmetry exists. 
When chiral symmetry is broken, the energy of the corner states can be shifted from the zero energy, and even merged into the bulk spectrum.
However, since the nontrivial value of $\nu_{\rm 2D}$ is determined only by the parity eigenvalues, the insulator without chiral symmetry cannot be adiabatically connected to a trivial insulator, which implies that the insulator without chiral symmetry still possesses nontrivial topology that is distinct from a trivial insulator.
In fact, the 2D topological system with/without chiral symmetry is called the second-order/third-order topological insulator (SOTI/TOTI)
~\cite{cualuguaru2019higher, okuma2019topological}.

The nontrivial band topology of the inversion symmetric HOTIs is manifested in filling anomaly, which is described as follows~\cite{song2017d, benalcazar2019quantization}.
In the chiral symmetric system, the number of occupied and unoccupied states are the same: $N_{\rm occ}=N_{\rm unocc}$.
For the inversion symmetric SOTI, there are two zero-mode corner states connected by the inversion symmetry.
However, the global geometry of the SOTI cannot satisfy the inversion symmetry when the system is half-filled, since only one of the corner states should be occupied (see Fig. 1(e), for example).
This phenomenon is called filling anomaly, which still holds even if chiral symmetry is broken, since the energy of the in-gapped states connected by the inversion symmetry are the same
~\cite{song2017d, benalcazar2019quantization}.

The low energy properties of 2D TOTI and its relation with 2D SOTI are given as follows. The low energy Dirac Hamiltonian for the boundary states of a 2D first-order TI can be written as $H_{edge}=-iR^{-1} \tilde{\sigma}_z\partial_{\theta}$, which describes the gapless chiral edge mode at the boundary. We use $\tilde{\sigma}$ to describe the symmetry representation for the surface states. For simplicity, we assume a disk-shaped geometry of the system with the radius $R$ and the corresponding polar angle $\theta$. For the inversion symmetric SOTIs, on the other hand, one can introduce mass term $m_1(\theta)\tilde{\sigma}_x$ which satisfies the constraint $m_1(\theta+\pi)=-m_1 (\theta)$ where inversion $P=\tilde{\sigma}_z$.
Thus, the edge Hamiltonian becomes
\begin{align}
H_{edge}=-iR^{-1}\tilde{\sigma}_z\partial_{\theta}+\tilde{m}_1(\theta) \tilde{\sigma}_x,
\end{align}
which is obviously chiral symmetric. Due to the constraint above, the mass gap is closed at two points at least, which indicates two corner states related by the inversion symmetry.

In the case of a 2D TOTI, chiral symmetry is broken so that two mass terms $m_1(\theta)\tilde{\sigma}_x$ and $m_2(\theta)\tilde{\sigma}_y$ are allowed, where $m_{1,2}(\theta+\pi)=-m_{1,2}(\theta)$. Thus, the surface spectrum is gapped in general. Nevertheless, it is still topological in the sense that mass winding number $w$ defined below is non-zero. 
\begin{align}
w=\int d\theta \frac{1}{2\pi} \partial_{\theta} \tan^{-1}\left(\frac{m_2(\theta)}{m_1(\theta)}\right)
\end{align}
This non-zero winding number is proportional to the charge accumulation at the boundary, which directly reflects the filling anomaly~\cite{wieder2018axion, hwang2019fragile}.

\section{Classification of Axion insulators by crystalline symmetries.}

In the main text, we mentioned that space-time orientation reversing symmetries quantize magnetoelectric polarizability $P_3$, and $H_{\text{3D}}(\bm{k})$ describes a 3D axion insulator.
The relevant symmetries of $H(k_x,k_y,\theta=0,~\pi)$ are $P$, $T$, $M_z$, $C_{2z}T$, $C_{4z}P$ and $C_{4z}T$.
The quantized $P_3$ of each 3D axion insulator can be described by a pumping process of 2D invariants that are defined at two different symmetry invariant planes with $k_z=0$ and $\pi$. 
Among the symmetries, $T$ and $M_z$ symmetric axion insulators are described by a pumping process of topological phases with the first order band topology (QSHI and mirror Chern insulators) and trivial phases.
For the other symmetries, the fractional charge at the order parameter vortex is described by a pumping process of a higher order topological insulator (HOTI) and a trivial insulator.
Here, we explicitly describe the low energy Hamiltonian and its physical property for each symmetry class.

\begin{figure}[h]
\centering
\includegraphics[width=8.5 cm]{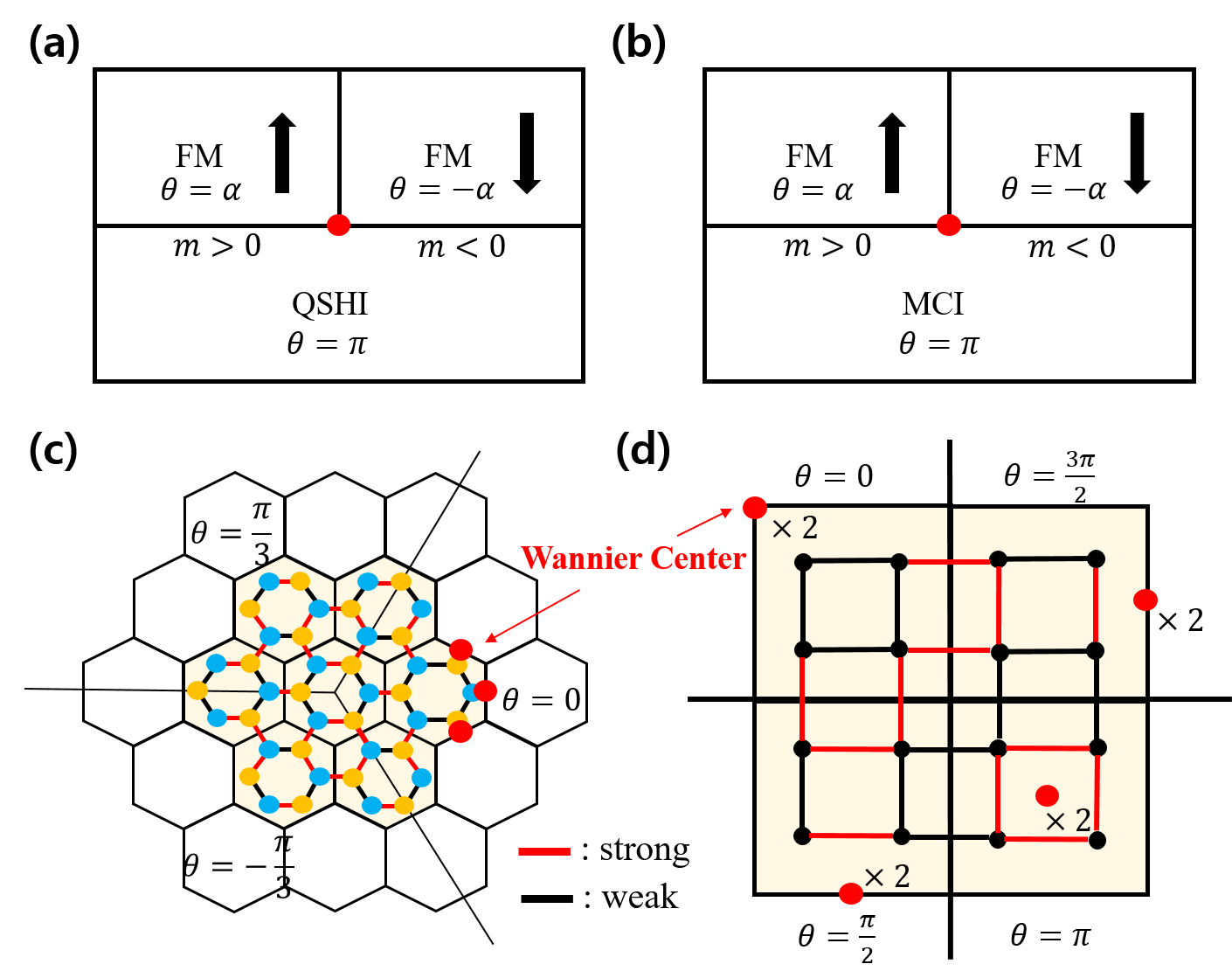}
\caption{(Color online) (a) A heterostructure where fractional charges are bound at the junction between a QSHI ($\theta=0$) and two ferromagnetic insulators (FM, $\theta=\pm\alpha$) with opposite in-plane magnetization. Since two FMs introduce surface mass terms with opposite sign, a zero-mode state is localized at the junction. (b) Similar heterostructure for a mirror Chern insulator.
(c) A $\mathbb{Z}_{3}$ vortex of Kekule textured graphene where three domains with lattice structure corresponding to $\theta=0$, $\pi/3$, $-\pi/3$ meet at a junction. Due to the Wannier function centered at the unit cell boundary at $\theta=0$, the vortex geometry contains half-integer electrons.
(d) A $\mathbb{Z}_{4}$ vortex relevant to $C_{4z}$ symmetric insulators with a quantized quadrupole moment where four domains with lattice structure corresponding to $\theta=0$, $\pi/2$, $\pi$, and $3\pi/2$ meet at a junction. Due to the Wannier functions centered at the unit cell corner at $\theta=0$, half-integer electrons are bound to the vortex center. 
} \label{fig:heterostructure}
\end{figure}

{\it $\mathcal{T}$ symmetry.|} 3D time reversal invariant $Z_2$ TI is described by a pumping process between a QSHI and a trivial insulator defined at the $k_z=0$ and $\pi$ planes~\cite{fu2006time}. Let us note that SOC is crucial since 2D AI class is trivial according to the tenfold classification~\cite{ryu2010topological}.
The low energy Hamiltonian for a 2D QSHI is
\begin{align}
H(\textbf{k})=k_x\tau_x\sigma_x+k_y\tau_x\sigma_y+M\tau_z\sigma_{0},
\end{align}
where Pauli matrices $\tau$ and $\sigma$ represent orbital and spin degrees of freedom and time reversal symmetry is $T=\sigma_yK$ satisfying $TH(\textbf{k})T^{-1}=H(-\textbf{k})$. The corresponding gapped insulator should be either a QSHI ($M>0$) or a trivial insulator ($M<0$) depending on the sign of the mass $M$~\cite{bernevig2006quantum}.
The Hamiltonian with a vortex can be expressed as 
\begin{align}
H(k_x,k_y,\theta)&=k_x\tau_x\sigma_x+k_y\tau_x\sigma_y
\nonumber\\
&+\Delta \cos\theta \tau_z\sigma_0+\Delta \sin\theta \tau_x \sigma_z,
\end{align}
where $\tau_x\sigma_z$ anticommutes with other matrices and it is odd under time reversal symmetry, so that the vortex structure breaks time-reversal symmetry except at $\theta=0$ (a trivial insulator) and at $\theta=\pi$ (a QSHI).

As shown in Fig.~\ref{fig:heterostructure} (a), a vortex structure can also be realized at the junction where a QSHI ($\theta=\pi$) and two ferromagnetic insulators ($\theta= \pm\alpha$) meet.
The Hamiltonian describing the gapless boundary of the QSHI is
\begin{align}
H_{0}&=v_F\int dx(\psi_{R}^{\dagger}i\partial_x\psi_{R}-\psi_{L}^{\dagger}i\partial_x\psi_{L} ),\nonumber \\
&=v_F\int dx\Psi^{\dagger}i\tilde{\sigma}_z \partial_x\Psi,
\end{align}
where $\psi_{R/L}$ is the right/left moving states and $\Psi^{\dagger}=(\psi_R^{\dagger},\psi_L^{\dagger}).$
The ferromagnetic insulators introduce time-reversal breaking mass terms that open the band gap of the edge state. The resulting the surface Hamiltonian is $H=H_0+\int dx\Psi^{\dagger}(m_1\tilde{\sigma}_x+m_2\tilde{\sigma}_y)\Psi$.
Parametrizing $m_1=m\cos\phi$, $m_2 = m\sin\phi$, the charge density at the magnetic domain is expressed as $\rho=\frac{1}{2\pi}\partial_x\phi$~\cite{goldstone1981fractional}.
When the spin direction is reversed, the sign of the mass terms is also reversed ($\phi\rightarrow\phi+\pi$).
Thus, the mass gap is closed at the junction, indicating a zero-mode state and fractional charges bound at the vortex center $Q=\int \rho dx = 1/2$~\cite{qi2008fractional}.

{\it $\mathcal{M}_z$ symmetry.|} For $\mathcal{M}_z$ symmetry, a mirror Chern number $C_m$ can be defined on the $k_z=0$ and $k_z=\pi$ planes. $C_m$ is given by the difference of the Chern number $C$ of two sectors with different mirror eigenvalues. The difference of the mirror Chern numbers on these planes corresponds to $2P_3$, that is, $C_m(k_z=0)-C_m(k_z=\pi) \equiv 2P_3 \mod 2$~\cite{ahn2018higher}.
Let us consider the low energy Hamiltonian describing a 2D mirror Chern insulator (MCI).
\begin{align}
H(\textbf{k})=k_x\Gamma_1+k_y\Gamma_2+M\Gamma_3,
\end{align}
where $\Gamma_1=\tau_x\sigma_x$, $\Gamma_2=\tau_x\sigma_y$, $\Gamma_3=\tau_z$. The mirror operator $M_z=-i\Gamma_1\Gamma_2\Gamma_3=\tau_z\sigma_z$ satisfies $M_zH(\textbf{k})M_{z}^{-1}=H(\textbf{k})$.
Considering the projection operator $P_{\pm}=\frac{1}{2}(1\pm \tau_z\sigma_z)$, the Hamiltonian can be expressed as a direct product of two Hamiltonians that have opposite mirror eigenvalues: $H=P_{+}H\oplus P_{-}H$.
Depending on the mass sign, the Hamiltonian describes either a MCI ($M>0, H=H_{C=+1}\oplus H_{C=-1}$) or a trivial insulator ($M<0$)~\cite{teo2008surface}.
The Hamiltonian with a vortex is given by 
\begin{align}
H(\textbf{k},\theta)&=k_x\tau_x\sigma_x+k_y\tau_x\sigma_y
\nonumber\\
&+\Delta_{1} \cos\theta \tau_z\sigma_0+\Delta_{2} \sin\theta \tau_x \sigma_z,
\end{align}
where $\tau_x\sigma_z$ is odd under mirror symmetry, so that the vortex structure breaks mirror symmetry except at $\theta=0$ and $\pi$.
Similar to the case with $\mathcal{T}$ symmetry, a vortex structure is realizable at a junction where a MCI and two mirror symmetry broken insulators meet. As shown in the Fig.~\ref{fig:heterostructure} (b), in spin orbit coupled systems, two ferromagnetic insulators with anti-parallel in-plane magnetization break mirror symmetry, introducing surface mass terms with opposite sign, so that fractional charges can be localized at the vortex core. In spinless fermion systems, the ferromagnetic insulators can be replaced by two ferroelectric insulators with anti-parallel out-of-plane charge polarization.

{\it $\mathcal{P}$ symmetry.|} 
Except for the cases with $\mathcal{T}$ and $\mathcal{M}_z$ symmetries, the quantized magnetoelectric polarizability $P_3$ of the axion insulators associated with other symmetries can be described by the pumping process between a higher order topological insulator and a trivial insulator.
The inversion symmetric axion insulator can be described by the pumping process between a $P$ protected higher order TI and a trivial insulator. The corresponding low energy Dirac Hamiltonian can be written as
\begin{align}
H(k_x,k_y,\theta)=k_x\Gamma_1+k_y\Gamma_2+M\cos\theta\Gamma_3+M\sin\theta\Gamma_4,
\end{align} 
where we take $P=\Gamma_3=\tau_z$, $\Gamma_1=\tau_x$, $\Gamma_2=\tau_y\sigma_z$, and $\Gamma_4=\tau_y\sigma_y$. One can clearly see that $H(k_{x},k_{y},\theta)$ is $P$ symmetric only at $\theta=0,~\pi$. Since the coefficient of the $\Gamma_3$ term has the opposite sign at $\theta=0$ and $\theta=\pi$, the number of the occupied bands with the negative $P$ eigenvalue is also different by two. 
Considering that the $Z_{2}$ invariant $\nu_{2D}$ of a $P$ symmetric 2D insulator is expressed by the product of $P$ eigenvalues at time reversal invariant momenta (TRIM) $\Gamma_i$ as
\begin{align}
(-1)^{\nu_{2D}}=\prod_{i=1}^{4}(-1)^{[N_{\text{occ}}^{-}(\Gamma_i)/2]},
\end{align}
where $N_{\text{occ}}^{-}(\Gamma_i)$ is the number of the occupied states with the negative $P$ eigenvalues at the momentum $\Gamma_i$. For a 2D higher order TI, $\nu_{2D}\equiv 1\mod 2 $ and the corresponding topological invariant for the axion insulator is given by $\nu_{2D}(k_z=0)-\nu_{2D}(k_z=\pi)\equiv 2P_3 \mod 2$~\cite{wieder2018axion}.
The higher order topology of the $\nu_{2D}=1$ phase can be understood as follows. Using the symmetry representation of $\Gamma$ matrices, the above Dirac Hamiltonian at $\theta=0,~\pi$ can be viewed as two copies of Chern insulators with the opposite Chern numbers. After suitable regularization by addition terms quadratic in momentum, one can make $H(\theta=0)$ to have a pair of counter-propagating edge modes whereas $H(\theta=\pi)$ has no edge mode. Then by adding $P$ symmetric mass terms to $H(\theta=0)$, one can find corner charges at the location where the sign of the surface mass terms changes~\cite{khalaf2018higher}.

{\it $\mathcal{C}_{4z}\mathcal{P}$ symmetry.|} 
The $\mathcal{C}_{4z}\mathcal{P}$ invariant axion insulator can be described by the pumping process between two different $C_{4z}$ invariant insulators defined on the $k_z=0$ and $k_z=\pi$ planes~\cite{van2018higher}.
For a 2D $C_{4z}$ invariant insulator, the corner charge $Q_C$ is quantized as $Q_C= \frac{N_C}{4}~\mod 1$, where $N_C$ denotes the number of electrons whose Wannier center is located at the Wyckoff position C. Thus, the higher order band topology of $2D$ $C_{4z}$ invariant insulators is characterized by a $\mathbb{Z}_4$ invariant.
However, a $C_{4z}P$ symmetric axion insulator is characterized by a $\mathbb{Z}_2$ invariant, which detects the difference of the corner charges defined on the $k_z=0$ and $k_z=\pi$ planes, respectively~\cite{van2018higher}.
The $\mathbb{Z}_2$ nature arises because $C_{2z}$ symmetry exists on every $k_z$ plane so that the set of $C_{2z}$ eigenvalues on the $k_z=0$ and $k_z=\pi$ planes must be the same in an insulating phase. Thus, the change of $N_C$ must be an even integer such as 
\begin{align}
N_C(k_z=\pi)-N_C(k_z=0)=0~\text{or}~2~\mod4.
\end{align}

Now let us consider the following low energy Dirac Hamiltonian
\begin{align}
H(\textbf{k},\theta)=k_x\Gamma_1+k_y\Gamma_2+M\cos\theta\Gamma_3+M\sin\theta\Gamma_4,
\end{align}
where $\Gamma_1=\sigma_x$, $\Gamma_2=\sigma_y$, $\Gamma_3=\tau_z\sigma_z$, $\Gamma_4=\tau_y\sigma_z$, $\Gamma_5=\tau_x\sigma_z$, and
$C_{4z}=$diag$[1,i,-1,-i]=\tau_z\otimes
\begin{pmatrix} 
1 & 0 \\
0 & i 
\end{pmatrix}$.
Then from $C_4\Gamma_1C_4^{-1}=\Gamma_2$ and $C_4\Gamma_2C_4^{-1}=-\Gamma_1$, one can see the $C_{4z}$ invariance of the Hamiltonian at $\theta=0,~\pi$, $C_4H(\textbf{k},\theta=0,~\pi)C_4^{-1}=H(C_4\textbf{k},\theta=0,~\pi)$.

At $\theta=0,~\pi$, the wave functions of the occupied states are $[0,1,0,0]^T$ and $[0,0,1,0]^T$ when $M>0$, with the corresponding $C_{4z}$ eigenvalues, $i$ and $-1$, respectively.  On the other hand, when $M<0$, the wave functions of the occupied states are $[1,0,0,0]^T$ and $[0,0,0,1]^T$ with the corresponding $C_{4z}$ eigenvalues, $1$ and $-i$, respectively. One can clearly see that the $C_{2z}$ eigenvalues of the occupied states remains the same independent of the sign of $M$. In this case, $\Delta N_C=2 \mod 4$ as shown in Ref.~[\onlinecite{van2018higher}], and the sign reversal of $M$ describes the nontrivial $\mathbb{Z}_2$ invariant of the axion insulator.

\begin{figure}[t]
\centering
\includegraphics[width=8.5 cm]{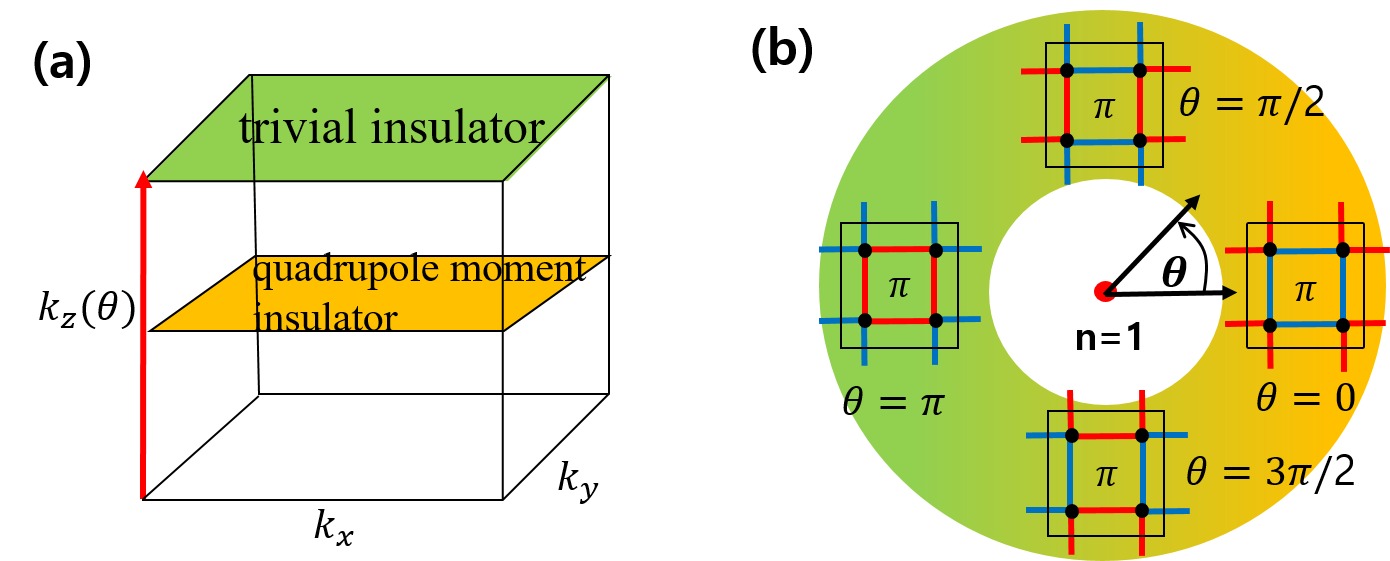}
\caption{(Color online) (a) Schematic figure describing the lattice structure around a vortex in a $C_{4z}$ invariant insulator. (b) Schematic figure describing the $\mathcal{C}_{4z}\mathcal{P}$ symmetric axion insulator in momentum space.  Here the $k_z(\theta)=0$ plane corresponds to the $2D$ insulator with a quantized quadrupole moment while  the $k_z(\theta)=\pi$ plane corresponds to a trivial insulator.
} \label{fig:quadrupole}
\end{figure}

On the other hand, when $\theta\neq0,~\pi$, $C_{4z}$ symmetry is broken while $C_{2z}$ is preserved. This means that around a vortex core, the system evolves continuously between two $C_{4z}$ invariant insulators while keeping the $C_{2z}$ symmetry. Fig.~\ref{fig:quadrupole} shows an example of the lattice structure modulation around the vortex. In real materials, since the $U(1)$ symmetry of the order parameter is reduced to a discrete $C_n$ symmetry due to the lattice potential. An example of $\mathbb{Z}_4$ vortex structure is shown in Fig.~\ref{fig:heterostructure} (d) in which fractional charges are localized at the junction between four domains with $\theta=0$, $\pi/4$, $\pi/2$, $3\pi/4$, respectively~\cite{chamon2008irrational,benalcazar2017electric}. Here the four domains are obtained due to four different ways choosing the unit cell for given lattice structure.

{\it $\mathcal{C}_{2z}\mathcal{T}$ symmetry.|} 
Let us first consider spinless fermion systems. The low energy Dirac Hamiltonian describing a $C_{2z}T$ invariant insulator can be written as
\begin{align}
H_{D}(\theta)=-i\Gamma_1\partial_1-i\Gamma_2\partial_2+M\cos\theta\Gamma_3+M\sin\theta\Gamma_4,
\end{align}
where $\Gamma_1=\tau_x$, $\Gamma_2=\sigma_y\tau_y$, $\Gamma_3=\tau_z$, $\Gamma_{4}=\sigma_z\tau_y$, $\Gamma_{5}=\sigma_{x}\tau_y$. Here we assume the following symmetry representations $C_{2z}=\Gamma_3=\tau_z$, $T=\Gamma_3 K$, $C_{2z}T=K$ where $K$ denotes complex conjugation. One can easily see that $H_{D}(\theta=0,~\pi)$ is $C_{2z}T$ invariant while $M\sin\theta\Gamma_4\neq0$ breaks $C_{2z}$ and $C_{2z}T$ while keeping $T$ invariance, and thus it can connect two $C_{2z}T$ invariant insulators with distinct topological properties.

The higher order nature of $C_{2z}T$ invariant insulators can be seen by following the similar idea as in the case of $C_{2z}$ symmetric systems. Namely, $H_{D}(\theta=0,~\pi)$  can be considered as two copies of quantum Hall insulators with opposite Chern numbers. Then after a suitable regularization, one can make $H_{D}(\theta=0)$ to have a pair of counter-propagating chiral edge modes whereas $H_{D}(\theta=0,~\pi)$ has no edge state. By adding $C_{2z}T$ symmetric mass terms, corner charges can be found at the domain wall of the surface mass terms.
Recently, it is shown that a $C_{2z}T$ invariant axion insulator, dubbed a 3D strong Stiefel Whitney insulator (SWI), can be described by using a pumping process between a 2D SWI and a trivial insulator with the corresponding $Z_2$ invariant $w_2=1$ and $w_2=0$, respectively. Also, it is shown that $w_2(k_z=0)-w_2(k_z=\pi) \equiv 2P_3 \mod 2$~\cite{wieder2018axion,ahn2018higher}.

In spinful fermion systems, one can use a different basis for symmetry representation. For instance, we can choose $C_{2z}=i\sigma_{y}$, $T=i\sigma_yK$, $C_{2z}T=K$, and also $\Gamma_1=\sigma_x$, $\Gamma_{2}=\sigma_z$, $\Gamma_3=\sigma_y\tau_y$, $\Gamma_4=\sigma_y\tau_x$, $\Gamma_5=\sigma_y\tau_z$ where $\sigma$ denotes the spin degrees of freedom.
One can see that $\Gamma_4$ and $\Gamma_5$ terms break $T$ symmetry while keeping $C_{2z}$. Hence the vortex structure breaks $T$ when $\theta\neq0,~\pi$ in spinful fermion systems, whereas it breaks $C_{2z}$ while keeping $T$ in spinless fermion systems.

{\it $\mathcal{C}_{4z}\mathcal{T}$ symmetry.|} 
The low energy Dirac Hamiltonian describing a QSHI protected by $C_{4z}$ and $T$ is
\begin{align}
H=-i\Gamma_1\partial_1-i\Gamma_2\partial_2+M\cos\theta\Gamma_3+M\sin\theta\Gamma_4,
\end{align}
where $\Gamma_1=\sigma_x\tau_x$, $\Gamma_2=\sigma_y\tau_x$, $\Gamma_3=\tau_z$, $\Gamma_{4}=\sigma_z\tau_x$, $\Gamma_{5}=\tau_y$, $C_4=\cos(\pi/4)+\sin(\pi/4)\Gamma_1\Gamma_2$ and $T=i\sigma_yK$. Using $C_{4z}\Gamma_1 C_{4z}^{-1}=-\Gamma_2$, $C_{4z}\Gamma_2 C_{4z}^{-1}=\Gamma_1$, one can easily check the invariance of the Hamiltonian under $C_{4z}$ and $T$. Also $\Gamma_{4}=\sigma_z\tau_x$ and $\Gamma_{5}=\tau_y$ break $T$ but satisfy $C_{4z}$. Hence a constant mass term $M\sin\theta\Gamma_4\neq0$ describing a ferromagnetic ordering with out-of-plane magnetization, breaks $T$ while keeping $C_{4z}$, and thus can connect two $C_{4z}T$ symmetric insulators with distinct topological properties. 

The higher order nature of the topological insulator protected by $C_{4z}T$ symmetry can be understood as follows. For convenience, let us consider a disk-shaped finite-size system with a circular boundary.
We assume that $M>0$ inside the insulator and $M<0$ outside the insulator. In polar coordinates, the Hamiltonian is expressed as 
\begin{align}
H=-i\Gamma_1(\theta)\partial_r-i\Gamma_2(\theta)r^{-1}\partial_{\theta}+M(r)\Gamma_3,
\end{align}
where $\Gamma_{1}(\theta)= \cos\theta \Gamma_{1} +\sin\theta\Gamma_{2}$ and $\Gamma_{2}=-\sin\theta \Gamma_{1} +\cos\theta \Gamma_{2} $.
Employing a projection operator $P(\theta)=$ $\frac{1}{2}(1+i\Gamma_1(\theta)\Gamma_3)$, we get a gapless surface Hamiltonian $H|_{r=R}=-iR^{-1}\tilde{\sigma_{z}}\partial_{\theta}$, which is a characteristic of a QSHI.
If $T$ symmetry is broken but non-local $C_4T$ symmetry is protected, surface mass terms are allowed: $H_{m}=m_4(\textbf{r})\Gamma_4+m_5(\textbf{r})\Gamma_5$. Projection to the surface gives $H_m|_{r=R}=m_4\tilde{\sigma_x}+m_5\tilde{\sigma_y}$.
Under $C_4T$ symmetry, mass terms change sign: $m_{i}(\textbf{r})=-m_{i}(C_4\textbf{r})$. Since two masses do not vanish simultaneously in general, the system does not exhibit anomalous corner states.
However, in the presence of chiral symmetry, for instance, $\Gamma_5 H \Gamma_5^{-1}=-H$, $m_5$ must vanish, so that $C_{4z}$ symmetric corner states appear.

The corresponding lattice Hamiltonian describing a $C_4T$ invariant axion insulator is proposed in Ref.~[\onlinecite{schindler2018higher}] as
\begin{align}
H(\textbf{k})&=(M+t\sum_{i}\cos k_i)\tau_z\sigma_0+\Delta_{1}\sum_{i}\sin k_i \tau_x \sigma_i \nonumber \\
&+\Delta_{2}(\cos k_x -\cos k_y) \tau_y\sigma_0 
\end{align}where $-3<M/t<-1$.
Between $k_z=0$ and $\pi$, mass term changes sign, which means that one of them describes a higher order topological insulator and the other is a trivial insulator.
Thus, the axion insulator exhibits a pumping process between a 2D chiral symmetric higher order topological insulator and a trivial insulator. 
Similar to the case with $\mathcal{T}$ symmetry, $(\mathcal{C}_{4z}\mathcal{T})^4=-1$ condition is crucial to define an axion insulator phase.
For instance, let us suppose that there is $\mathcal{C}_{4z}\mathcal{T}$ symmetric chiral hinge modes for a finite size system. On the top surface, there must be gapless modes that respect $\mathcal{C}_{4z}\mathcal{T}$ symmetry. However, gapless modes are not protected for spinless fermion systems. Whereas gapless modes are protected for spinful electron systems since $+i$ and $-i$ eigenvalues of $\mathcal{C}_{2z}$ symmetry are always paired due to $\mathcal{C}_{4z}\mathcal{T}$ symmetry.

\section{Field-theoretical description for a topological vortex}
Let us explain the connection between the topological spectral flow and charge accumulation on a vortex by exploiting field theoretical approach. Since the number of topological spectral flow lines is equivalent to the winding number of the vortex, it suffices to show the relation between the winding number and vortex charge.
The low energy Hamiltonian describing a vortex can be written as follows (we assume that the vortex is protected by inversion symmetry, without loss of generality)~\cite{chamon2008electron,ryu2009masses}.
\begin{align}
H= \Psi^{\dagger}[\alpha\cdot{\bf p}+\beta(\Delta_1-\gamma_5\Delta_2)+\mu]\Psi,
\end{align}
where $p=-i(\partial_x,\partial_y), \alpha=(\sigma_x \tau_z,\sigma_y\tau_z), \beta=\tau_x, \gamma_5=\tau_z, \Delta(r)=\Delta_1(r)+\Delta_2(r)=\Delta_0(r)e^{i\theta(r)}, \mu$ is the next-nearest interaction term that breaks chiral symmetry $S=\sigma_z\tau_z$, and inversion symmetry is represented as $\beta$, such that the Hamiltonian is invariant under inversion only at $\theta=0,\pi$. $\theta=0$ corresponds to a higher-order topological insulator (HOTI) phase, and $\theta=\pi$\ corresponds to a trivial insulator phase as in the main text.

Under a unitary transformation, the Hamiltonian can be written as $H=
\begin{pmatrix}
\mu & D \\
D^{\dagger} & \mu 
\end{pmatrix}$, 
where $D=i\sigma^i\partial_i+i\Delta_1+\sigma_3\Delta_2$, and $D^{\dagger}=i\sigma^i\partial_i-i\Delta_1+\sigma_3\Delta_2$~\cite{chamon2008electron}.
One can see that there is a bulk gap between $E=-m+\mu$ and $E=m+\mu$ $(m=|\Delta|)$. 
If the vortex has a winding number $n$, there are $n$ bound states between the gap. To prove this, we solve the Schrodinger equation.
\begin{align}
H\Psi_{E}=E\Psi_E, \Psi_{E}=(u_E,v_E)^{t}
\end{align}
First, let us find the solution where $E=\mu$, such that
\begin{align}
\begin{pmatrix}
0 & D \\
D^{\dagger} & 0 
\end{pmatrix}
\begin{pmatrix}
u_E \\
v_E  
\end{pmatrix}=0.
\end{align}
The equation is equivalent to $Dv_E=0, u_E=0$ or $D^{\dagger}u_E=0,v_E=0.$
According to the index theorem, the number of solutions $Dv_E=0$ and $D^{\dagger}u_E=0$ are determined by the winding number of vortex order parameter $\Delta_0(r)e^{i\theta}$:~\cite{weinberg1981index}
\begin{align}
n>0: \dim kerD^{\dagger} = 0,\quad \dim ker D = n,\nonumber\\
n<0: \dim kerD^{\dagger} = |n|,\quad \dim ker D = 0.
\end{align}
The index theorem tells us that the number of in-gapped bound states is equal to the winding number of the vortex.

Second, let us find the solution where $E\neq\mu$. Let $E=\mu+\epsilon$, then one can find that valence and conduction states are always paired ($E=\mu\pm\epsilon$) since $D^{\dagger}Du_E=\epsilon^2 u_E$. Thus, the number of occupied bound states is equal to the number of unccupied bound states.

In conclusion, as vortex order parameter winds the vortex core once, a nontrivial bound state is created. Especially in the presence of chiral symmetry, the nontrivial state is fixed at the zero energy, which gives an exact half electric charge localized at the vortex core.

\section{Wannier center and quantized Wannier charge in topological vortex}
\begin{figure}[t]
\centering
\includegraphics[width=8.5 cm]{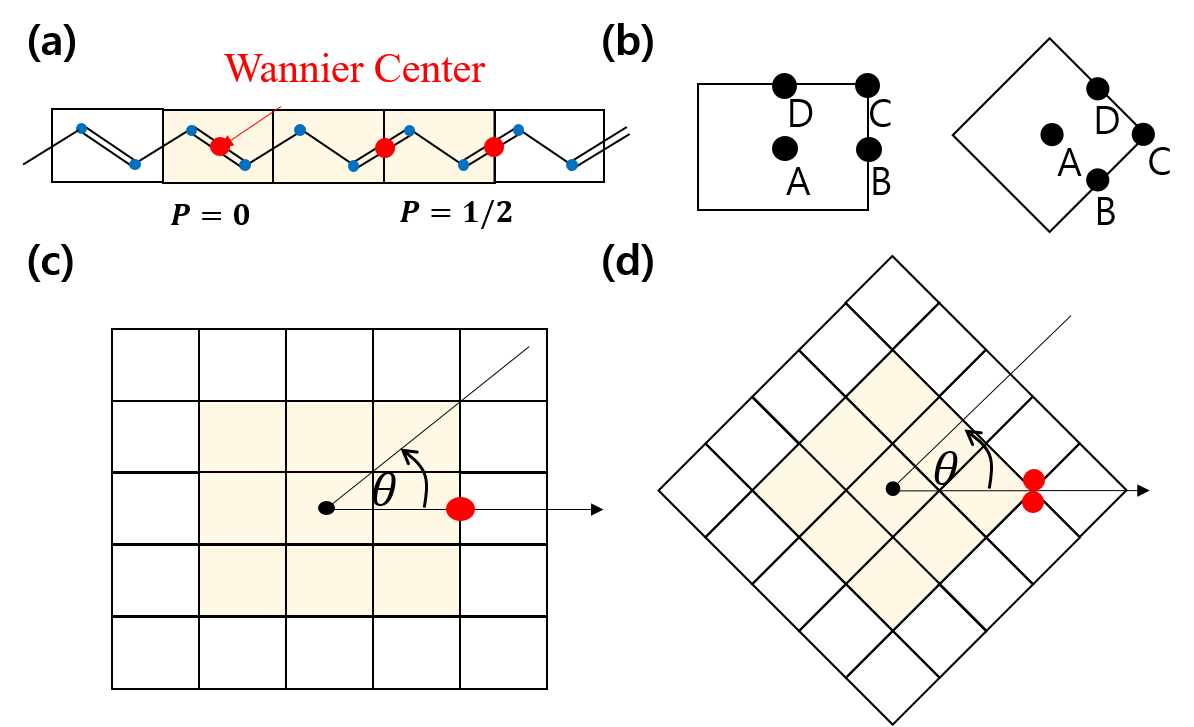}
\caption{(Color online) (a) Domain wall structure between trivial ($P_1=0$) and topological($P_1=1/2$) insulator. Since Wannier function is located at the unit cell boundary for $P_1=1/2$, the domain wall structure composed of integer unit cells (shaded region) contains half integer electrons. (b) Wyckoff positions of $P$ and $C_{4z}$ symmetric lattices. (c) Topological vortex protected by inversion symmetry. For $\theta=0$ (inversion symmetric HOTI), $N_B=N_C=N_D=1\mod 2$. Since Wannier function is located at the unit cell boundary at $\theta=0$, the charge bound to the vortex is quantized to half-integer.
(d) Topological vortex protected by $C_{4z}$ symmetry. Since $N_C(\theta=0)=2\mod 4$ ($C_{4z}$ symmetric HOTI), the charge bound to the vortex is $\frac{1}{4}(2N_C(\theta=0)+2N_c(\theta=\pi)) =\frac{1}{2} \mod 1$.
} \label{fig:Vortex}
\end{figure}

Here we use symmetric Wannier function description to show that fractional Wannier charge bound to a topological vortex is quantized even without chiral or particle-hole symmetry.
Let us first consider a 1D domain wall system of the SSH Hamiltonian. The value of polarization should be quantized to $0$ or $1/2 \mod 1$ in the presence of inversion symmetry. When $P_1=0$, Wannier function is located at the center of the unit cell. In the case of $P_1=1/2$, on the other hand, Wannier function is located at the unit cell boundary. Fig.~\ref{fig:Vortex} (a) describes the domain wall structure of the SSH Hamiltonian between trivial and topological phases. 
Here we assume that chiral symmetry is broken due to, for instance, the next nearest neighbor hopping. 
On the left-hand side($P_1=0$), the Wannier center is located at the unit cell center. On the other side($P_1=1/2$), the Wannier center is located at the unit cell boundary, where two adjacent cells share an electron: the electron gives a half charge contribution to each unit cell. Therefore, the domain structure composed of unit cells contains half-integer electrons due to the electron located at the unit cell boundary on the right-hand side. Namely, irrespective of chiral or particle-hole symmetry, a vortex always hosts a quantized half-integral Wannier charge as long as the relevant symmetry is preserved. Although the quantized Wannier charge is not identical to the physical electric charge localized at a domain wall when chiral symmetry is broken, the half-quantization of the Wannier charge at a domain wall indicates the fact that the two insulators meeting at the domain carry distinct quantized topological invariants and therefore cannot be smoothly connected to each other in the presence of inversion symmetry.

Similarly, the existence of half-quantized charge bound to a topological vortex can be explained by the Wannier function description. 
Except for $T$ and $M_z$, there are four symmetries $P$ (or $C_{2z}$), $C_{2z}T, C_{4z},$ and $C_{4z}T$ that are relevant to the vortices with higher order topology. Since $T$ is local-in-space symmetry, it does not change the position of Wannier center. Thus, to prove that half-integer charge is bound to a topological vortex, we only have to consider $P$ (or $C_{2z}$) and $C_{4z}$ symmetries.

{$P$ (or $C_{2z}$).|}
Inversion symmetric insulator has four Wyckoff positions A, B, C, and D which are invariant under inversion up to lattice translation vector (Fig.~\ref{fig:Vortex} (a)).
We define the number of symmetric Wannier functions centered at each Wyckoff position as $N_A,N_B,N_C,$ and $N_D$.
For inversion symmetric HOTI, polarization is zero: 
\begin{align}
p_x=\sum_{i}X_i=\frac{1}{2}(N_B+N_C)=0 \mod 1,\\
p_y=\sum_{i}Y_i=\frac{1}{2}(N_C+N_D)=0 \mod 1, 
\end{align}
where $X_i$ and $Y_i$ denote $x$ and $y$ coordinates of $i$th Wannier function, respectively.
On the other hand, quadrupole moment is defined as
\begin{align}
q_{xy}=\sum_{i}X_iY_i=\frac{1}{4}N_C=\frac{1}{4} \mod \frac{1}{2}.
\end{align}
Let us note that $N_C$ is well defined modulo 2 since $N_C$ can be changed only by multiples of two keeping inversion symmetry. 
Taken altogether, $N_B=N_C=N_D=1\mod2$.
For a trivial insulator, on the other hand, quadrupole moment $q_{xy}=\frac{1}{4}N_C=0 \mod \frac{1}{2}$.
Thus, $N_B=N_C=N_D=0\mod2$.
Let us consider a vortex structure $P H(k_x,k_y,\theta) P^{-1}= H(-k_x,-k_y,-\theta)$ where $\theta=0$ $(\pi)$ corresponds to inversion symmetric HOTI (trivial insulator) (Fig.~\ref{fig:Vortex} (c)).
Since an odd number of Wannier functions are located at the unit-cell boundary for $\theta=0$, the vortex geometry composed of integer unit-cells contains a half-integer charge.

{\it $C_{4z}$.|}
$C_{4z}$ symmetric insulator has two Wyckoff positions A and C which are invariant under $C_{4z}$ operation and two Wyckoff positions B and D which are invariant under $C_{2z}$ operation up to lattice translation vector (Fig.~\ref{fig:Vortex} (b)).
The vortex structure satisfies $C_{4z} H(k_x,k_y,\theta) C_{4z}^{-1}= H(k_x',k_y',-\theta)$, followed by $C_{2z} H(k_x,k_y,\theta) C_{2z}^{-1}= H(-k_x,-k_y,\theta)$.
Since we are not interested in inversion symmetry protected topological phase, $N_B=N_C=N_D=0\mod2$ for any $\theta$.
As shown earlier, in the case of $C_{4z}$ symmetric HOTI, $N_C= 2 \mod 4$.
Also, $N_C=0\mod4$ for trivial insulator.
Note that $N_C$ is well defined modulo 4 since $N_C$ can be changed by multiples of four keeping $C_4$ symmetry.
The charge bound to the vortex is, therefore, $\frac{1}{4}[N_C(\theta=0)+N_c(\theta=\pi)] =\frac{1}{2} \mod 1$, for the electron located at the Wyckoff position $C$ gives $1/4$ charge contribution to each unit cell (Fig.~\ref{fig:Vortex} (d)).


\end{document}